\documentclass[twocolumn]{aastex63}
\usepackage[version=4]{mhchem}
\usepackage{graphicx, multirow}
\usepackage{array}

\received{XXX}
\revised{YYY}
\accepted{ZZZ}
\submitjournal{PSJ}

\shortauthors{Currie et al.}

\graphicspath{{./}{figures/}}

\begin{document}

\title{There's more to life in reflected light: Simulating the detectability of a range of molecules for high-contrast, high-resolution observations of non-transiting terrestrial exoplanets}

\correspondingauthor{Miles Currie}
\email{miles.h.currie@nasa.gov}

\author[0000-0003-3429-4142]{Miles H. Currie}
\affiliation{Department of Astronomy and Astrobiology Program, University of Washington, Box 351580, Seattle, Washington 98195, USA}
\affiliation{NASA Nexus for Exoplanet System Science, Virtual Planetary Laboratory Team, Box 351580, University of Washington, Seattle, Washington 98195, USA}
\affiliation{NASA Goddard Space Flight Center, Greenbelt, MD 20771, USA}

\author[0000-0002-1386-1710]{Victoria S. Meadows}
\affiliation{Department of Astronomy and Astrobiology Program, University of Washington, Box 351580, Seattle, Washington 98195, USA}
\affiliation{NASA Nexus for Exoplanet System Science, Virtual Planetary Laboratory Team, Box 351580, University of Washington, Seattle, Washington 98195, USA}
\affiliation{Astrobiology Center, 2-21-1 Osawa, Mitaka, Tokyo 181-8588, Japan}

\begin{abstract}
The upcoming extremely large telescopes will provide the first opportunity to search for signs of habitability and life on non-transiting terrestrial exoplanets using high-contrast, high-resolution instrumentation. However, the suite of atmospheric gases in terrestrial exoplanet environments that are accessible to ground-based reflected light observations has not been thoroughly explored. In this work, we use an upgraded Extremely Large Telescope (ELT) detectability pipeline to simulate the detectability of gases that can serve as habitability markers, potential biosignatures, and false positive discriminants in the atmospheres of Earth-sized and sub-Neptune planets. 
We calculate molecular detectability for five photochemically self-consistent atmosphere types, including the modern and Archean Earth, uninhabited biosignature ``false positive'' environments, and a sub-Neptune, over a grid of observational configurations for non-transiting targets within 10pc of Earth. 
For the most accessible nearby target, Proxima Centauri b, our results suggest that we may be able to rule out a sub-Neptune atmosphere in as little as a single hour of observing, and two biosignature disequilibrium pairs (\ce{O2}/\ce{CH4} and \ce{CO2}/\ce{CH4}) may be accessible in $\sim 10$ hours for the most optimistic scenario. It may also be possible to discriminate uninhabited worlds, and rule out biosignature false positives by identifying contextual indicators (\ce{CO} and \ce{H2O}) of abiotic \ce{O2} and/or \ce{CH4} buildup on similar timescales. In the near term, ELT reflected light observations will likely allow us to characterize multiple nearby terrestrial atmospheres, and ultimately search for signs of habitability and life. 
\end{abstract}


\section{Introduction}\label{sec:intro}

In just a few decades, the scientific community has made major advances toward discovering and characterizing extrasolar planets, and is now poised to search for signs of habitability and life in the atmospheres of rocky, terrestrial-sized exoplanets. JWST has enabled the first observations of the surfaces and/or atmospheres of the Earth-sized TRAPPIST-1 planets: while initial data on TRAPPIST-1 b and c are consistent with airless worlds \citep{Greene2023, Zieba2023}, the presence of atmospheres has not been conclusively ruled out \citep{Zieba2023, Ih2023,Lincowski2023, Ducrot2024}. However, JWST is not our only near term tool for characterizing terrestrial planets. Within the next decade, the next-generation extremely large ground-based telescopes (ELTs) \citep[e.g.][]{Szentgyorgyi2014-xj, Marconi2022, Palle2023, Mawet2019} will begin to come online, and may also be powerful tools for terrestrial exoplanet characterization \citep[e.g.][]{Snellen2013, Rodler2014, Lopez-Morales2019, Currie2023, Hardegree-Ullman2023}, including for non-transiting planetary targets \citep[e.g.][]{Lovis2017-rr, Wang2017, Hawker2019-lt, Zhang2023, Vaughan2024}.

While transiting terrestrial exoplanets will be accessible to both JWST and the ELTs \citep[e.g.][]{Lustig-Yaeger2019-bk, Currie2023}, the ELTs will also have the sensitivity to characterize the larger sample of nearby non-transiting habitable zone terrestrial exoplanets \citep[e.g.][]{Snellen2015-tu, Hawker2019-lt}. Because the edge-on geometry required for a transit is relatively rare, there are far more non-transiting planets, and observational access to them provides more---and closer---targets for characterization studies. Among the known nearby ($< 5$ pc away) exoplanets, a handful are likely rocky and within the habitable zone of their host star, including Proxima Centauri b \citep[1.3 pc distant, M6V host][]{Anglada-Escude2016-ai}, GJ 1061 d \citep[3.67 pc, M5.5V host][]{Dreizler2020-ii}, Teegarden’s Star c \citep[3.83 pc, M7V host][]{Zechmeister2019-pa}, and GJ 1002 b and c \citep[4.85 pc, M5.5V host][]{Suarez_Mascareno2022-hd}. Due to its proximity to Earth, Proxima Centauri b has been the most well studied in theoretical ELT detectability simulations, which assume its atmosphere to be Earth-like \citep[e.g.][]{Lovis2017-rr, Wang2017, Hawker2019-lt, Zhang2023, Vaughan2024}; however, the true mass, and corresponding composition of the Proxima Centauri b atmosphere, is still unknown and there is at least a 10\% probability that the planet is not of terrestrial mass, but is a sub-Neptune \citep{Bixel2017}.

To characterize non-transiting planets using reflected stellar light, the high planet-to-star contrast ratio must be overcome by combining high contrast imaging (HCI) and high-resolution spectroscopy (HRS), a technique originally proposed by \citet{sparks_imaging_2002}.  HCI systems use a coronagraph to suppress the significant diffracted starlight that may otherwise overwhelm the planet signal \citep[e.g.][]{Guyon2006-ez}. Current state-of-the-art ground-based systems are able to achieve a suppression level of $\sim10^{-4}$ \citep{Macintosh2014, Beuzit2008, Jovanovic2015}, allowing for the detection of some gas giant or brown dwarf objects in images observed from the ground \citep[e.g.][]{Wagner2016}. HRS alone can also be used to detect planets that are $\sim 10^{-4} \times$ fainter than their host stars by taking advantage of the differences in the radial velocity and spectral absorption lines between the planet and host star \citep{Birkby2018}. However the sensitivity to planets is enhanced by pairing HCI with a high-resolution spectrograph (HRS) \citep{Snellen2015-tu} to allow for a combined (multiplicative) starlight suppression level of up to $\sim 10^{-8}$ or more with current capabilities \citep{Wang2017}, and theoretically up to $\sim 10^{-10}$ or better for future instruments \citep{Snellen2015-tu}. 

Combining HCI and HRS systems, hereafter referred to as high dispersion coronagraphy (HDC), is the focus of current and future efforts to develop instrumentation capable of characterizing terrestrial exoplanet atmospheres \citep{Snellen2015-tu}. First proposed by \citet{sparks_imaging_2002, Riaud2007} and later refined to explore prospects for Earth-like targets by \citet{Snellen2015-tu}, this technique has been demonstrated on currently operating telescopes.  For example, the CRIRES instrument \citep{Kaeufl2004} has been successfully paired with the MACAO adaptive optics system \citep{Arsenault2003} on the ESO Very Large Telescope (VLT), enabling the detection and characterization of giant planet atmospheres \citep[e.g.][]{Snellen2014-ny, Schwarz2016, Hoeijmakers2018, Bryan2018}. However, to detect terrestrial planet atmospheres, a new class of extreme adaptive optics systems must be developed to achieve contrasts better than $\sim10^{-7}$. Successful demonstrations of extreme HDC systems in a lab environment \citep[e.g.][]{Mawet2017} have led to proposals that couple the SPHERE high-contrast imager with the ESPRESSO high-resolution instrument on the VLT \citep{Lovis2017-rr}, to produce the RISTRETTO HDC instrument concept \citep{Lovis2022}. RISTRETTO's primary science goal is to  detect the reflected light from Proxima Centauri b at a contrast of about $10^{-7}$, and it will be a foundational proof-of-concept instrument for the development of similar systems on the ELTs. These include the SCAO-IFU mode on the ANDES instrument of the European ELT (E-ELT) \citep{Marconi2022, Palle2023}, the ELT-PCS instrument \citep{Kasper2021}, the Planetary Systems Imager \citep{Jensen-Clem2021} and Second Earth Imager \citep{Matsuo2010, Matsuo2012} on the Thirty Meter Telescope (TMT), and GmagAOx paired with G-CLEF on the Giant Magellan Telescope (GMT) \citep{Males2022}.

In advance of the next generation of instrumentation and telescopes, theoretical studies have made important contributions in predicting the performance of current and future ground-based facilities for Earth-like exoplanet studies. Because \ce{O2} is the main byproduct of oxygenic photosynthesis on Earth and a potential biosignature gas \citep{Hitchcock1967-ky, Meadows2018-gh}, significant emphasis has been placed on estimating the detectability of \ce{O2} in Earth-like atmospheres for both transit transmission (i.e. HRS-only) \citep[e.g.][]{Rodler2014, Lopez-Morales2019, Snellen2013, Currie2023, Hardegree-Ullman2023} and reflected light (HDC) \citep[e.g.][]{Hawker2019-lt, Lovis2017-rr, Wang2017, Zhang2023, Vaughan2024} observations. While \ce{O2} detection will likely be challenging for transiting targets \cite[e.g.][]{Hardegree-Ullman2023, Currie2023}, \ce{O2} may be more accessible in reflected light observations of non-transiting exoplanets, because the targets will be closer and detections may require less exposure time \citep[e.g.][]{Hawker2019-lt, Lovis2017-rr, Lovis2022}.

However, there are known abiotic (planetary) sources of \ce{O2}, and its use as a biosignature will depend on our ability to gain further environmental context to determine its source and/or abundance in the atmosphere \citep{Meadows2017}. To enhance the interpretation of \ce{O2} as a biosignature gas, a simultaneous detection of \ce{CH4}, the canonical \ce{O2}/\ce{CH4} biosignature disequilibrium pair \citep{Hitchcock1967-ky, Meadows2017}, could indicate an additional biosphere of methanogenic organisms. This biosignature requires biogenic fluxes of both \ce{O2} and \ce{CH4} gases to explain their continued presence in the Earth's atmosphere \citep{Hitchcock1967-ky}. Also, while known photochemistry can produce up to $\sim$percent amounts of \ce{O2} \citep{Domagal-Goldman2014-oh, Tian2014-fs, Harman2015-oc, Gao2015-ck, Hu2012-jr, Harman2018}, the 21\% \ce{O2} abundance in the modern Earth atmosphere is difficult to achieve abiotically, although may be possible in the case of atmospheres with a low non-condensible gas component \citep{Wordsworth2014-be}.  %biological flux required to maintain this concentration, and the simultaneous saturation of \ce{O2} sinks over time in buried oxidized iron and sulfur compounds \citep{Lyons2014}.
%gaining further environmental context can also help to rule out abiotic \ce{O2} in so-called ``false-positive'' environments.  
However, very large abundances of abiotic \ce{O2} could be generated via the photolysis of \ce{H2O} in scenarios where a planet may be undergoing ocean evaporation \citep{Luger2015, Meadows2018-yx}.
%in addition to serpentinization that can generate some atmospheric \ce{CH4}. 
In this scenario, \ce{O2} can build up multi-bar atmospheres \citep{Luger2015,Schaefer2016}, and the false positive discriminant is significant \ce{O2}--\ce{O2} collisionally-induced absorption features in the planetary transmission spectrum \citep{Lustig-Yaeger2019-bk} that may also suppress one or more \ce{O2} absorption bands in high-resolution observations \citep{Leung2020}. 

We can also consider other gas disequilibria as an indicator for life: \ce{CO2} and \ce{CH4} may be an alternative pair of biosignature disequilibrium gases that were first proposed for Earth's \ce{O2}-poor ancient Archean atmosphere \citep{Krissansen-Totton2018-tj, Meadows2023}. Similar to the \ce{O2}/\ce{CH4} pair, a high abundance of \ce{CH4} in the presence of significant \ce{CO2} requires a high flux of \ce{CH4} to maintain a detectable abundance in the atmosphere against atmospheric oxidation via \ce{CO2}.  If the \ce{CH4} flux can be quantified, it can be assessed as to whether it is more or less likely to be due to less productive abiotic processes like serpentinization, when water runs over ultramafic rock \citep{Guzman-Marmolejo2013, Etiope2013-fg}, or more productive biological processes such as methanogenesis \citep{Krissansen-Totton2018-tj, Meadows2023}. Although originally proposed for the Archean \citep{Krissansen-Totton2016-bq}, the \ce{CO2}/\ce{CH4} biosignature pair may also be detectable in transmission for M dwarf planets with a modern Earth photosynthetic biosphere and methanogenesis \citep{Meadows2023}.  This makes the \ce{CO2}/\ce{CH4} disequilibrium biosignature significantly more persistent throughout Earth's history when compared to the \ce{O2}/\ce{CH4} pair, which may have been detectable only within the last 1 billion years \citep{Planavsky2018}, and a prime  target for observational studies of terrestrial exoplanet atmospheres \citep{Meadows2023}. However, in lifeless environments such as the prebiotic Earth, significant volcanism could potentially generate high abundances of \ce{CH4} that may be detectable in atmospheres with significant \ce{CO2}, which can be a false positive for the \ce{CO2}/\ce{CH4} biosignature pair \citep{Krissansen-Totton2016-bq,Wogan2020, Thompson2022, Meadows2023}. In this prebiotic Earth scenario, the volcanism would also generate large abundances of \ce{CO} in that atmosphere, which would be the discriminant for this false positive, as \ce{CO} is not prevalent in a biological Earth-like scenario \citep{Krissansen-Totton2016-bq}.  In summary, while the \ce{O2}/\ce{CH4} and \ce{CO2}/\ce{CH4} disequilibrium pairs can be robust indicators of life, there are also ``false positive'' scenarios where these gases can build up abiotically, and gaining environmental context will be key to discriminating inhabited from lifeless worlds.

A handful of recent studies modeling high-resolution spectra of terrestrial planets go beyond \ce{O2} detection alone to investigate the detectability of other molecules that can help provide environmental context, and find that a range of molecular features may be detectable with the ELTs in transmission \citep{Currie2023}, and potentially in reflected light \citep{Wang2017, Lovis2017-rr, Zhang2023, Vaughan2024}. In particular, \cite{Currie2023} suggested that the \ce{CH4}/\ce{CO2} biosignature pair \citep{Krissansen-Totton2016-bq, Krissansen-Totton2018-tj, Meadows2023}, may be highly detectable in ELT observations of transiting Earth-like planets. \citet{Wang2017, Lovis2017-rr, Zhang2023} predict that \ce{H2O}, \ce{O2}, \ce{CO2}, and \ce{CH4} may be detectable with future instrumentation for modern Earth-like atmospheres, however considerations for the detectability of these molecules in the astrobiological context of both inhabited and false positive abiotic environments for non-transiting terrestrial planets observed in reflected light has not yet been explored.

To prepare for future ELT observations of non-transiting terrestrial exoplanet atmospheres and better understand biosignature detection and interpretation using the ELTs, we investigate the detectability of a wide range of molecular features in simulated spectra of photochemically self-consistent terrestrial and sub-Neptune atmospheres orbiting M dwarf stars.  In Section~\ref{sec:methods}, we describe our methods including generating synthetic HDC observations and detecting molecular features via cross-correlation. We present our results in Section~\ref{sec:results}, discuss the meaning of our results in an astrobiological context in Section~\ref{sec:discussion}, and conclude in  Section~\ref{sec:conclusions}.

\section{Methods}\label{sec:methods}
In this work, we modify the high-resolution, cross-correlation detectability pipeline of \citet{Currie2023} to include treatment of non-transiting planets orbiting M dwarf host stars. As input to the detectability pipeline, we use high-resolution ($R\sim 100,000$) spectra of climatically and photochemically self-consistent terrestrial and sub-Neptune planet environments. We simulate observations of these atmospheres for a range of observational and instrumental configurations, and compute cross-correlation functions to search for a range of molecular absorption bands for 1000 realizations of noise. We finally estimate the detection significance for each molecular absorption band as a function of exposure time.  Below, we first describe how we generate synthetic planetary spectra for the planets considered here, and then describe the SPECTR observation simulator, including its upgrades, cross-correlation mechanism, and noise sources.

\subsection{Planetary Spectra}\label{sec:planetaryspectra}
For the terrestrial planetary atmospheres considered in this study, we use a similar suite of previously generated photochemically self-consistent temperature and molecular mixing ratio profiles used in \citet{Currie2023}, with the addition of a \ce{CO2}/\ce{CH4} biosignature false positive environment (Prebiotic Earth)\citep{Meadows2023}, and a sub-Neptune type atmosphere from \citet{Charnay2015}. These atmospheric profiles are passed to the 1D radiative transfer model SMART \citep{Meadows1996-fp,Crisp1997-fn} to calculate high resolution reflected light spectra, which are subsequently used as input to our cross-correlation detectability pipeline \citep{Currie2023}.

We consider four distinct classes of terrestrial planet atmospheres in this study, and a sub-Neptune world for comparison.  The inhabited environments we consider include the pre-industrial Earth-like (modern Earth with a photosynthetic biosphere and without anthropogenic fluxes) and Archean Earth-like atmospheres orbiting around M dwarf stars  5 giga-years in age from Davis et al. (in prep), both of which have active biospheres. These atmospheres were also used in \citet{Currie2023} to test molecular detectability for transiting terrestrial planets, and atmospheric profiles from Davis et al. 2025 (in prep) are plotted in Figures 3~and~4 of \citet{Currie2023}. These planets are placed in orbit around M2V, M3V, M4V, M6V, or M8V host stars such that they receive 67\% of the instellation that Earth receives (corresponding to an orbit on the outer edge of the habitable zone) and driven to climate-photochemical equilibrium. 

We note that our pre-industrial and Archean Earth-like atmospheres include 10\% \ce{CO2} (Davis et al. 2025 (in prep)), which our climate model indicates is required to maintain a globally-averaged surface temperature close to present Earth’s (275--318 K, see Table~\ref{tab:atmosphere_composition}), and well above the freezing point of water \citep{Meadows2018-yx, Meadows2023}. This 10\% \ce{CO2} abundance is reasonable if we assume that the surface temperature of these planets is buffered by an active carbonate--silicate cycle which maintains planetary habitability as the star brightens \citep{Walker1981-wj}, or in this case, for planets that are further from the star than Earth’s equivalent distance. If the carbonate-silicate cycle is active, then we would expect that planets further from the star would have progressively higher atmospheric \ce{CO2} fractions, with planets at the outer edge habitable zone limit requiring several bars of \ce{CO2} to maintain a habitable surface temperature \citep{Kopparapu2013-wm}.  We leverage this diversity of atmospheres and stellar hosts to understand how M dwarf host star type may affect the detectability of spectral features in  non-transiting terrestrial planets. The compositions of our atmosphere environments are listed in Table~\ref{tab:atmosphere_composition}. We assume that all planets have a similar distribution of land, water, and ice as the modern Earth, however we remove the contribution of vegetation from the Archean, ocean loss, and prebiotic Earth cases, resulting in surface albedos of 0.112 and 0.124 for cases with and without vegetation, respectively.

\begin{deluxetable*}{cccc}\label{tab:atmosphere_composition}
    \tablecaption{Composition of terrestrial atmospheric cases}
    \tablehead{
        \colhead{Atmosphere template} & \colhead{Surface} & \colhead{Atmospheric gases} & \colhead{Surface Temperature [K]}
    }
    \startdata
    Pre-industrial Earth & Earth Composite & \begin{tabular}[t]{@{}p{5.5cm}@{}} 68\% \ce{N2}, 21\% \ce{O2}, 10\% \ce{CO2}, 0.2\% \ce{H2O}, 700-800 ppm \ce{CH4}, trace \ce{O3}, \ce{CO} \end{tabular} & 275 \\
    Archean Earth & Earth no vegetation & \begin{tabular}[t]{@{}p{5.5cm}@{}} 78\% \ce{N2}, 10\% \ce{CO2}, 0.2\% \ce{H2O}, 100-400 ppm \ce{CH4}, 2 ppm \ce{C2H6}, trace \ce{O2}, \ce{O3}, \ce{CO} \end{tabular} & 275 \\
    Ocean Loss & Earth no vegetation  & \begin{tabular}[t]{@{}p{5.5cm}@{}} 95\% \ce{O2}, 4\% \ce{N2}, 0.5\% \ce{CO2}, 0.3\% \ce{H2O}, trace \ce{O3}, \ce{CO}, \ce{NO2} \end{tabular} & 318 \\
    Prebiotic Earth$^*$& Earth no vegetation & \begin{tabular}[t]{@{}p{5.5cm}@{}} 89\% \ce{N2}, 1.9\%~\ce{CO2}, 4.4\% \ce{CO}, 4.8\% \ce{H2}, 14~ppm \ce{CH4}, trace \ce{O2}, \ce{H2O} \end{tabular} & 262 \\
    \enddata
    \tablenotetext{*}{low redox volcanism, 200$\times$ enhanced}
\end{deluxetable*}

We also consider habitable, but  uninhabited, biosignature false positive atmospheres that can build up potentially detectable levels of gases that are also generated by life. In this work, we consider a false positive scenario for both the pre-industrial and Archean Earth like atmospheres.  For our pre-industrial photosynthetic biosphere false positive, we use the high-\ce{O2}, ocean retaining planet from \citet{Meadows2018-gh}. A planet that undergoes significant ocean loss  when its younger star is more luminous, could build up tens to 1000s of bars of abiotic \ce{O2} via the photolysis of evaporated \ce{H2O} and the subsequent loss of hydrogen to space \citep{Luger2015,Schaefer2016,Meadows2018-gh, gialluca2024implications}. For a false positive for the \ce{CO2}/\ce{CH4} biosignature we use the modeled version of a prebiotic Earth with enhanced volcanic outgassing from a more reducing mantle from \citet{Meadows2023}. In this scenario, both \ce{CO2} and abiotic \ce{CH4} are volcanically outgassed, and build up in the atmosphere \citep{Krissansen-Totton2018-tj, Meadows2023}.  

\begin{figure}
    \centering
    \includegraphics{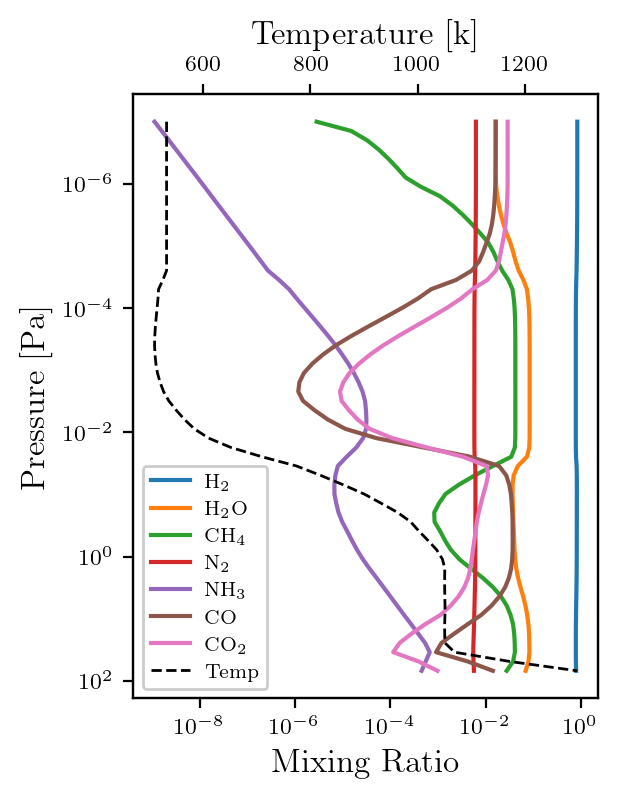}
    \caption{Mixing ratios for the major molecular species in the sub-Neptune atmosphere, and the atmospheric temperature profile \citep{Charnay2015}. }
    \label{fig:spaghetti_subneptune}
\end{figure}

For comparison, we also consider a sub-Neptune type atmosphere in thermochemical equilibrium, as calculated by \citet{Charnay2015} for GJ 1214 b. Figure~\ref{fig:spaghetti_subneptune} shows the molecular and temperature vertical  profiles for the sub-Neptune atmosphere. The GJ 1214 b atmosphere is a significantly reducing atmosphere, containing high levels of \ce{H2} and \ce{NH3}, as well as molecules that are found in terrestrial atmospheres, including \ce{CH4}, \ce{CO2}, and \ce{H2O}. The molecular bands we study are listed in Table~\ref{tab:bands}.  We note that this planet is closer to its star than the habitable zone, with an equilibrium temperature close to 500K (Figure \ref{fig:spaghetti_subneptune})

\begin{deluxetable}{c|cc}\label{tab:bands}
\tablecaption{Molecular bands explored for the atmospheres in this study}
\tablewidth{0pt}
\tablehead{
\colhead{} & \colhead{Molecule} &  \colhead{Bands [$\mu m$]}
}
\startdata
\multirow{6}{*}{\rotatebox[origin=c]{90}{Terrestrial}} & O$_2$ & 0.69, 0.76, 1.27 \\
 & CH$_4$ & 0.89, 1.1, 1.3, 1.6  \\
 & CO$_2$ & 1.59, 2.0 \\
 & H$_2$O & 0.94, 1.1, 1.3 \\
 & O$_3$ & 0.63, 0.65, 3.2 \\
 & CO & 1.55, 2.3 \\
\hline
\multirow{5}{*}{\rotatebox[origin=c]{90}{Sub-Neptune}} & CH$_4$ & 0.89, 1.1, 1.3, 1.6  \\
 & CO$_2$ & 1.59, 2.0 \\
 & CO & 1.55, 2.3 \\
 & H$_2$O & 0.94, 1.1, 1.3 \\
 & NH$_3$ & 1.43, 1.9, 2.15 \\
 \enddata
\end{deluxetable}

Our assumed planet and host star properties are given in Table~\ref{tab:planet_star}, and planetary contrast ratios are given in Table~\ref{tab:contrast_ratios}.  For the M dwarf terrestrial planets, we use the same stellar spectra as in \citet{Currie2023}, and that paper also includes a more detailed description and figures related to the planetary and stellar parameters. For each spectral type, we use publicly available synthetic high-resolution stellar spectra for the stars GJ832, GJ436, GJ876, Proxima Centauri, and TRAPPIST-1 (\citet{Peacock2019-bz, Peacock2019-ol, Peacock2020-cc}, Davis et al. in prep.), serving as representative examples for M2V, M3V, M4V, M6V, and M8V dwarf stars, respectively.  To simulate habitable zone targets, the planet is placed at an orbital radius such that it receives 0.66 times the irradiance that Earth receives, which is approximately the radiation received by Proxima Centauri b \citep{Anglada-Escude2016-ai}.  In the GJ 1214 b case, we assume the planet has a radius of $\sim 2.68$ Earth radii, and a semi major axis of 0.0143 au \citep{Charbonneau2009}. 

\begin{deluxetable*}{c|ccccc}\label{tab:planet_star}
    \tablecaption{Host star and planetary companion properties for the terrestrial atmospheres in this study}
    \tablehead{
    \colhead{} & \colhead{M2V host} & \colhead{M3V host} &\colhead{M4V host} &\colhead{M6V host} &\colhead{M8V host}
    }
    \startdata
    R$_{*}$ & 0.499$^a$ & 0.464$^b$ & 0.376$^c$ & 0.141$^d$ & 0.114$^e$ \\
    m$_I$ (5pc) & 6.2 & 6.8 & 7.1 & 10.7 & 12.2 \\
    m$_J$ (5pc) & 5.0 & 5.4 & 7.1 & 10.7 & 12.2 \\
    \hline
    R$_{p}$ & 1 & 1 & 1 & 1 & 1 \\
    P [day] & 48 & 33 & 29 & 11 & 6.1 \\
    a [AU] & 0.24 & 0.19 & 0.16 & 0.041 & 0.027 \\
%    atmospheres & PIE, ARE & PIE, ARE & PIE, ARE, \ce{CO2 phot} & PIE, ARE, \ce{H2O} phot. & PIE, ARE \\
    \enddata
\tablerefs{
$^a$\citet{Houdebine2010-cu},
$^b$\citet{Torres2007-cz},
$^c$\citet{Von_Braun2014-xs},
$^d$\citet{Bonfils2005-pb},
$^e$\citet{Filippazzo2015-wv}}
\end{deluxetable*}

\begin{deluxetable*}{c|c|cccccc}\label{tab:contrast_ratios}
    \tablecaption{Planet-to-star flux ratios for the atmospheres in this study in the photometric bands V, R, I, J, H, and K. The values listed are for a planet at quadrature.}
    \tablehead{\colhead{Atmosphere Type} & \colhead{Host Star} & \colhead{V-band} & \colhead{R-band} & \colhead{I-band} & \colhead{J-band} & \colhead{H-band} & \colhead{K-band}  }
    \startdata
\multirow{5}{*}{\begin{tabular}{c}Pre-industrial\\
                                      Earth-like
                \end{tabular}} & M2V &  $2.16 \times 10^{-9}$ & $1.97 \times 10^{-9}$ & $2.81 \times 10^{-9}$ & $1.43 \times 10^{-9}$ & $4.24 \times 10^{-10}$ & $1.16 \times 10^{-10}$ \\
& M3V &  $3.53 \times 10^{-9}$ & $3.30 \times 10^{-9}$ & $4.63 \times 10^{-9}$ & $2.36 \times 10^{-9}$ & $6.92 \times 10^{-10}$ & $1.88 \times 10^{-10}$ \\
& M4V &  $4.87 \times 10^{-9}$ & $4.48 \times 10^{-9}$ & $6.18 \times 10^{-9}$ & $3.06 \times 10^{-9}$ & $8.82 \times 10^{-10}$ & $2.34 \times 10^{-10}$ \\
& M6V &  $7.64 \times 10^{-8}$ & $7.89 \times 10^{-8}$ & $1.01 \times 10^{-7}$ & $5.12 \times 10^{-8}$ & $1.40 \times 10^{-8}$ & $3.74 \times 10^{-9}$ \\
& M8V &  $1.69 \times 10^{-7}$ & $1.79 \times 10^{-7}$ & $2.08 \times 10^{-7}$ & $1.10 \times 10^{-7}$ & $3.09 \times 10^{-8}$ & $9.03 \times 10^{-9}$ \\
\hline
\multirow{5}{*}{\begin{tabular}{c}Archean\\
                                      Earth-like
                \end{tabular}} & M2V &  $2.25 \times 10^{-9}$ & $2.11 \times 10^{-9}$ & $3.06 \times 10^{-9}$ & $1.66 \times 10^{-9}$ & $5.53 \times 10^{-10}$ & $1.62 \times 10^{-10}$ \\
& M3V &  $3.70 \times 10^{-9}$ & $3.56 \times 10^{-9}$ & $5.02 \times 10^{-9}$ & $2.80 \times 10^{-9}$ & $1.00 \times 10^{-9}$ & $2.96 \times 10^{-10}$ \\
& M4V &  $5.06 \times 10^{-9}$ & $4.74 \times 10^{-9}$ & $6.57 \times 10^{-9}$ & $3.38 \times 10^{-9}$ & $1.11 \times 10^{-9}$ & $3.23 \times 10^{-10}$ \\
& M6V &  $8.10 \times 10^{-8}$ & $8.44 \times 10^{-8}$ & $1.07 \times 10^{-7}$ & $5.87 \times 10^{-8}$ & $2.00 \times 10^{-8}$ & $5.90 \times 10^{-9}$ \\
& M8V &  $1.81 \times 10^{-7}$ & $1.93 \times 10^{-7}$ & $2.21 \times 10^{-7}$ & $1.26 \times 10^{-7}$ & $5.10 \times 10^{-8}$ & $1.62 \times 10^{-8}$ \\
\hline
\multirow{1}{*}{Ocean-loss} & M6V &  $2.53 \times 10^{-7}$ & $1.42 \times 10^{-7}$ & $1.05 \times 10^{-7}$ & $2.54 \times 10^{-8}$ & $2.23 \times 10^{-8}$ & $1.43 \times 10^{-8}$ \\
\hline
\multirow{1}{*}{Prebiotic Earth-like}  & M6V &  $8.55 \times 10^{-8}$ & $8.96 \times 10^{-8}$ & $1.15 \times 10^{-7}$ & $8.20 \times 10^{-8}$ & $4.41 \times 10^{-8}$ & $1.65 \times 10^{-8}$ \\
\enddata
\end{deluxetable*}

After compiling our library of terrestrial planet atmospheres, we calculate high-resolution reflectance spectra of our planet atmospheres using LBLABC/SMART, a 1D, line-by-line radiative transfer model \citep{Stamnes1988,Meadows1996-fp, Crisp1997-fn}. LBLABC calculates the absorption coefficients for each molecule in the atmosphere, then SMART calculates the optical properties of the atmosphere at a given wavelength and atmospheric layer. In each layer of the atmosphere, SMART calculates extinction due to vibrational and rotational transitions, and collisionally-induced absorption for each absorbing gas. It also calculates the effects of aerosols, Rayleigh scattering, and wavelength-dependent surface albedo. SMART outputs top-of-atmosphere planetary radiances and transmission spectra, and has been validated for Earth in reflected light at low-resolution in \citet{Robinson2011-os}, and for Earth in transmission at high-resolution in \citet{Lustig-Yaeger2022}.  We calculated each spectrum at a resolution of $R=1,000,000$ and convolved with a Gaussian profile to achieve a spectral resolution of $R=100,000$ for this study.

\subsection{Synthesizing Observations with SPECTR}
We convert our high-resolution planetary spectra into synthetic observations using the Spectral Planetary ELT Calculator for Terrestrial Retrieval (SPECTR) detectability pipeline, which is described in detail in \citep{Currie2023}. SPECTR is a comprehensive, sophisticated noise model for ground-based telescopes and instrumentation that is based on the \texttt{coronagraph} noise model for space-based telescopes \citep{Robinson2016-lw, Lustig2019coronagraph}. The original \texttt{coronagraph} model can be used to simulate noise and realistic observations using future high-contrast instrumentation on space-based telescopes, with the goal of detecting and characterizing exoplanets primarily with low spectral resolution. SPECTR, on the other hand, takes high resolution spectra as input, simulates realistic ground-based observations, including noise from the thermal state of the observatory and instrument/detector properties including photon noise, read noise, and dark current, and additionally includes the effects of the Earth's telluric contamination, and finally applies a cross-correlation analysis to determine the detectability of spectral features 
 \citep{Currie2023}.    %See \citet{Currie2023} for a full description of SPECTR and its capabilities. 

\subsubsection{Upgrading SPECTR to simulate reflected light observations}
SPECTR was initially developed to simulate transit transmission spectroscopy \citep{Currie2023}, and in this work, we have developed new features to upgrade SPECTR with the capability to make detectability estimates for non-transiting planets in reflected light. 
%These upgrades include the addition of a simplistic stellar light suppression model for high-contrast observational modes, and the simulation of the effects of inclination and phase dependencies, including appropriate RV calculations. 
We now provide a simplistic stellar suppression model for high-contrast imaging by allowing the user to multiply the stellar light by a specified contrast value, $C$, which parameterizes the level of stellar suppression for a given observational configuration, as in e.g. \citet{Wang2017, Hawker2019-lt, Zhang2023}. %The contrast level is specified by the user when running a SPECTR simulation by simply adjusting the contrast variable. 
To support reflected light simulations, the code has been modified to calculate the planet's observable illumination fraction as a function of viewing geometry, taking into account system orientation and the planet's phase. The detectability of the molecules in the planetary atmosphere is also dependent on the its radial velocity (Section~\ref{sec:rv}). 
%Measuring the spectrum of a non-transiting planet in reflected light differs from the measurement of a transiting planet: 
This upgrade is necessary because, unlike a transiting planet, a non-transiting planet can be observed at the range of  positions in its orbital path that is not limited by inner working angle constraints. 
%Therefore, the planet's detectability is highly dependent on not only the host star's RV, but also the system's inclination and the planet's phase. 
SPECTR  calculates the planet's RV according to the observing geometry using the following equation:
\begin{equation}
    \mathrm{RV}_p = K_p (\cos(\phi + \omega) + e\cos(w)), 
\end{equation}
where $\mathrm{RV}_p$ is the planetary RV, $K_p$ is the planetary semi-amplitude, $\phi$ is the planetary phase, $\omega$ is the argument of periastron, and $e$ is the orbital eccentricity. The planetary semi-amplitude encodes inclination ($i$), orbital radius ($a$), and orbital period ($P$) and is given by 
\begin{equation}
    K_p = 2 \pi \frac{a  \sin(i)}{P\sqrt{1 - e^2}}, 
\end{equation}
where the orbital radius (semi-major axis, $a$) is given by:
\begin{equation}\label{eq:a}
    a = \left[\frac{G (m_s + m_p)P^2}{4\pi^2}\right]^{1/3}.
\end{equation}
In Equation~\ref{eq:a}, G is the gravitational constant and $m_s$ and $m_p$ are the mass of the star and planet, respectively.
For this work, we assume the planet is on a circular orbit and the orbital plane is 60$^\circ$ inclined from face-on; however, we included functionality in SPECTR to model eccentric orbits for future work.

\subsubsection{Observational assumptions and IWA limitations}
Because we are testing the relative detectability of molecular features, and not the efficiency of instrumentation or telluric removal algorithms, we make several simplifications to our observational configuration. We assume that all observations occur at the planet's maximum separation (quadrature). For telluric line removal, we assume that the Earth's telluric contribution to the spectrum is known (Section~\ref{sec:tellurics}) and can thus be removed from the data via a cross-correlation analysis between the template spectrum and the known telluric transmittance (Section~\ref{sec:telluric_removal}).  Additionally, we assume noise due to the planetary and stellar speckle fluxes as well as readout noise and dark current on the detector. We note that these assumptions result in optimistic detectability estimates, and quoted observation times should in general be interpreted as lower limits.

\begin{figure}
    \centering
    \includegraphics{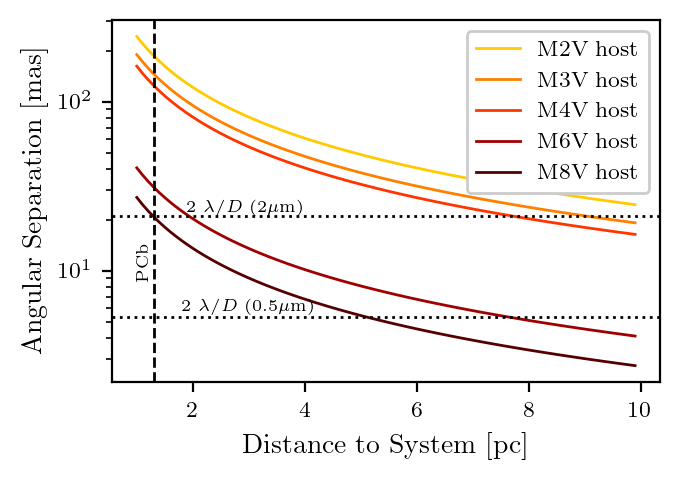}
    \caption{Angular separation at quadrature as a function of distance to the system for planets receiving 67\% of Earth's instellation in the habitable zones of a range of host star types. The habitable zones of earlier type M-dwarfs reside at larger orbital radii than later type systems, thus are less affected by observational configurations limited by coronagraphic IWA. However, achieving the necessary contrast for planets orbiting earlier type hosts will inherently be more challenging, and early type systems will indeed require higher integration times in general if we adopt constant instrumental contrast, as shown later in this work. Each planet is in its respective system’s habitable zone, and thus all planets receive a similar instellation to achieve a habitable climate, regardless of spectral type.  
    %Therefore, the semi-major axis squared dependence due to the habitable zones being farther away for an early M dwarf is countered by the definition of the habitable zone, and a planet in the habitable zone of an early star receives and reflects similar amounts of light to a planet in the habitable zone of a late star.  
    }
    \label{fig:ang_sep}
\end{figure}

Where appropriate, we also consider whether the angular separation of the planet from the star is outside of our assumed inner working angle of 2$\lambda/D$ for the imaging instrumentation. 
%We also do not take into account whether the angular separation of the planet and star is sufficient to avoid inner working angle (IWA) constraints in the imaging instrumentation, but for the nearby targets discussed in this work, this is unlikely to be an issue for the later type systems and shorter wavelength bands we target.
One of the top-level requirements of the RISTRETTO instrument concept on the VLT is that it have an IWA of  $\sim 2 \lambda / D$ \citep{Lovis2022}. While the final specifications of the ELT instruments are not yet known, if they are designed to achieve a similar IWA requirement as RISTRETTO the typical angular separation limit as a function of wavelength will be $\sim5-20$ mas in the 0.5--2.0 $\mu$m wavelength range for the E-ELT's 39 m diameter. The nearby (1.3 pc) Proxima Centauri b target, which orbits an M6V star, would have the entire  0.5--2.0 $\mu$m wavelength range accessible near quadrature (Figure~\ref{fig:ang_sep}), but planets in the habitable zones of late-type M dwarf targets farther away (e.g. GJ 1061 d and GJ 1002 b/c at $\sim4$pc) may become challenging because the longer wavelength molecular bands will not be accessible. In particular, the 1.6 $\mu$m \ce{CO2} and \ce{CH4} bands will begin to become inaccessible at a distance of $\sim$ 2.5 and 1.6 pc for the M6V and M8V hosts, respectively (Figure~\ref{fig:ang_sep}). However, we may not be as limited by IWA constraints when observing planets orbiting early type M dwarfs, whose habitable zones are farther from the star--- a planet in the habitable zone of an M4V star may be accessible in the 0.5--2.0 $\mu$m wavelength range out to $\sim7$pc, and out to $\sim10$pc for an M2V planet. We plot the angular separations for our planets as a function of distance in Figure~\ref{fig:ang_sep} for comparison with future instrument design specifications. Observational cases that fall inside the assumed IWA are discussed further in Sections~\ref{sec:results} and~\ref{sec:discussion}.

\subsubsection{Observational Configuration, Noise Sources, and Telluric Contamination}\label{sec:tellurics}
To simulate sources of noise and signal attenuation associated with the sky and background, SPECTR interfaces with the Cerro Paranal Advanced Sky Model \citep[SkyCalc][]{Noll2012-br, Jones2013-ex} to  incorporate signal extinction due to telluric spectral contamination into our simulated observations.  For all telescope configurations considered here, we use Paranal as our observatory site for all three ELTs simulated (see below), as a proxy for the similar dry, high altitude sites identified for the ELTs. We assume  a precipitable water vapor column of 3.5 mm, and do not include scattered moonlight.

\begin{figure}
    \centering
    \includegraphics[width=0.45\textwidth]{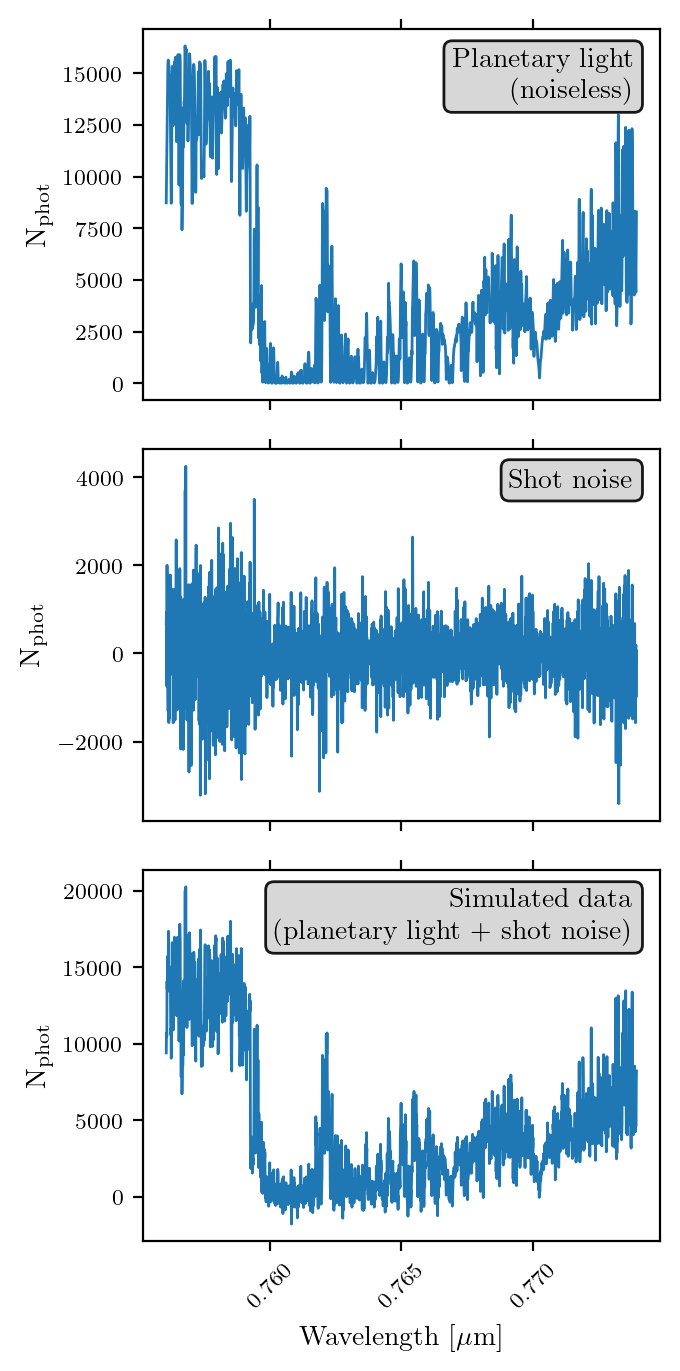}
    \caption{Example of simulated data for a 1000 hour observation of the \ce{O2} A-band (lower panel) for an Earth-like target orbiting an M6V host star 1.3 pc away using the 39 m ELT. The simulated data is constructed with two components: the light reflected from the planet and the shot noise, plotted in the upper and middle panels, respectively. The shot noise is composed of planetary, stellar, and background sources (see Section~\ref{sec:tellurics}), and we assume a 10\% total throughput. }
    \label{fig:photons}
\end{figure}

We simulate high spectral resolution observations for European ELT (E-ELT), Thirty Meter Telescope (TMT), and Giant Magellan Telescope (GMT) configurations, with collecting areas of and 978 m$^2$, 707 m$^2$, and 368 m$^2$, respectively. These collecting areas correspond to a 39 m E-ELT with a central hole to accommodate its secondary mirror\footnote{\url{https://elt.eso.org/about/facts/}}, a 30 m diameter TMT, and the total expected collecting area of all mirror segments of the GMT\footnote{\url{https://giantmagellan.org/explore-the-design/}}. The ELTs will be equipped with high-resolution spectrometers capable of R $\sim$ 100,000 in visible and/or near-infrared wavelengths with estimated throughputs of 10\%, typical dark current values of 0.0002 \ce{e-}/pix/s and 0.015 \ce{e-}/pix/s for the visible and NIR, respectively, and read noise of 3 \ce{e-}/pix \citep[GMT]{Szentgyorgyi2014-xj} \citep[E-ELT]{Marconi2022, Palle2023} \citep[TMT]{Mawet2019}. We use these values for our E-ELT, TMT, and GMT configurations for this study, and assume 10 detector pixels per resolution element, which is on the order of magnitude for the default value of the E-ELT ANDES Exposure Time Calculator\footnote{\url{http://tirgo.arcetri.inaf.it/nicoletta/etc_andes_sn_com.html}}, and a maximum exposure time per frame of one hour to limit the RV-induced spectral smearing on the detector as in \citet{Lovis2017-rr}. 
The observation is synthesized using the following equation:
\begin{equation}\label{eq:data}
    \mathrm{data} = F_p + \mathcal{N}(0,1)\sqrt{F_p + F_s + F_n + \mathrm{RON}^2}, 
\end{equation}
where $F_p$ and $F_s$ are the planet and stellar fluxes observed by the telescope, and $F_n$ encompasses the sky background and dark current contributions outlined above, and RON is the readout noise. Additionally, $\mathcal{N}(0,1)$ represents an array of random draws, with length equal to the length of the flux and noise arrays, from a normal distribution with zero mean and a standard deviation of one.   We note that our simulations implicitly assume that the contaminating stellar spectrum, $F_s$ can be perfectly removed from the observation, such that only the planetary flux $F_p$ and noise remain. See \citet{Currie2023} for a more detailed description of how SPECTR handles $F_n$. Examples of our simulated data, planetary light, and noise components are plotted in Figure~\ref{fig:photons}.

\begin{figure}
    \centering
    \includegraphics{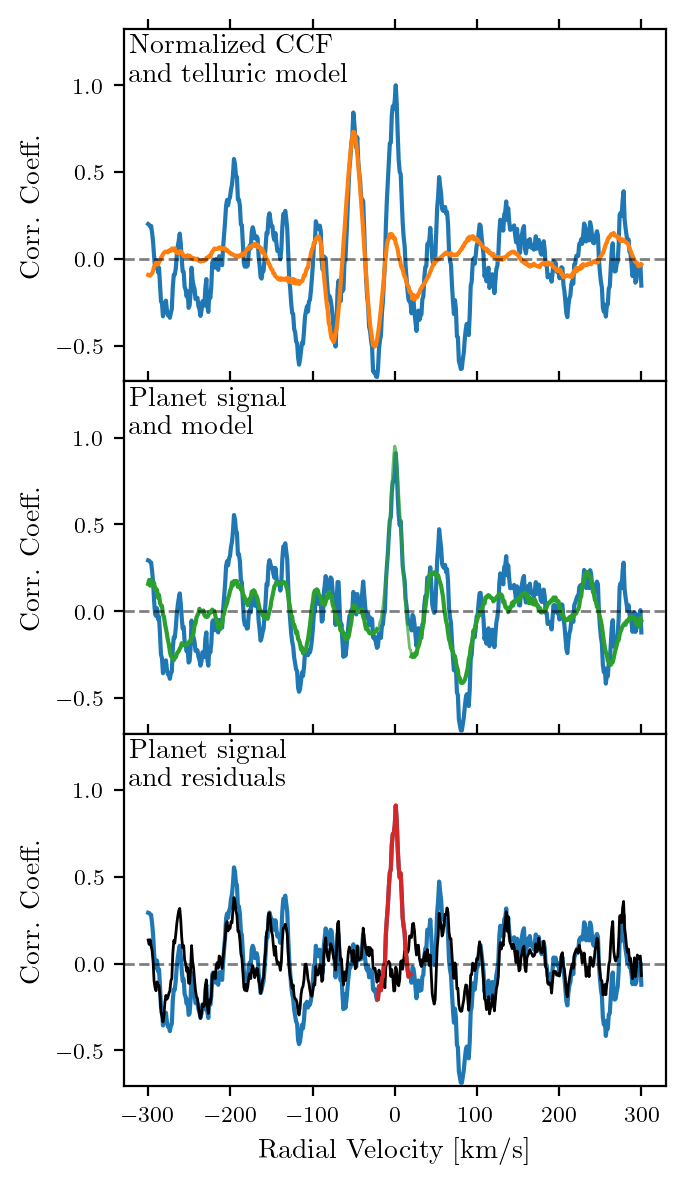}
    \caption{Stages of cross-correlation analysis, including telluric line removal (upper), alias detrending (middle), and S/N estimation (lower). The data cross-correlation function (CCF) is shown in blue, and the model telluric and planet CCFs are in orange and green, respectively. The upper panel shows the telluric CCF fit to the raw data CCF, which is subtracted off to remove the tellurics from the observation. The middle panel shows the telluric-subtracted data CCF (blue) as well as the planetary model CCF (green), which is used for detrending the aliasing patterns. The lower panel shows the detrended CCF residuals in black and the planetary signal in red, as well as the telluric-subtracted data CCF curve (blue) for comparison. The S/N is calculated by dividing the maximum signal coefficient by the standard deviation of the detrended CCF residuals.}
    \label{fig:ccf}
\end{figure}

\subsection{Cross-correlation Analysis}
To estimate the detectability of the molecular absorption bands in our simulated data, we employ a cross-correlation technique similar to that described in \citet{Brogi2016}.  This technique includes sensitivity to line locations, line shapes, relative line depths, and robustness against small perturbations to the radial velocity due to stellar or planetary processes when applied to real data. We cross-correlate our simulated reflected light spectra with a model template of the molecular absorption band based on the techniques tested and used by similar studies \citep[e.g.][]{Snellen2010, Snellen2013, Rodler2014, Brogi2016, Serindag2019, Lopez-Morales2019, Spring2022, Currie2023}, and report detections as the significance at the expected signal location in the resulting cross-correlation functions. To simulate the impact of possible molecular contamination from the stellar spectrum, we assume that the reflected stellar light in the planetary flux is not divided out.

To reduce contamination from other spectral features within the observed wavelength range during the cross-correlation, we use a model template spectrum containing only the molecular species of interest, even though the simulated data will include features from all species modeled in the atmosphere.  This reduces contamination in the cross-correlation function and extracted S/N measurement. The Doppler velocity of the template spectrum is allowed to vary over $\pm300$ km/s on an evenly spaced grid of 301 elements.

\subsubsection{Telluric line removal}\label{sec:telluric_removal}
Because we make the simplistic assumption that all observations occur at quadrature, it is straightforward to directly remove telluric line contamination from the raw cross-correlation function of the data itself. This is achieved by first cross-correlating our model template spectrum with the telluric line SkyCalc model assumed for the observations, resulting in a telluric model cross-correlation function. We then fit this telluric model CCF to the raw data CCF, and subtract the telluric contribution from the data CCF. This process is illustrated in the top panel of Figure~\ref{fig:ccf}, and the residual telluric-free data CCF is plotted in blue in the middle panel. 

However, we note that in a realistic observing scenario, more advanced techniques will be employed to remove telluric contamination because the planet will not be stationary for the duration of the observation when it moves along a finite portion of its orbit. These methods will separate the planetary signal from the stationary telluric lines as the planet moves along its orbital path using PCA techniques \citep[e.g.][]{Brogi2018-dn}, or by modeling the atmospheric contamination itself \citep[e.g.][]{Allart2017-ja, Smette2015-jl, Bertaux2014}. More recent advancements show that removing tellurics on a sub-pixel level may also be possible with increased spectral resolution and knowledge of the stellar spectrum \citep{Cheverall2024}. In this work, we have opted to simplify telluric contamination and removal to focus on the relative detectability of molecular features, and do not explore these advanced techniques.

\subsubsection{Alias Detrending}
The off-peak values of cross-correlation functions include predictable patterns due to the structure of the molecular band that can introduce systematic noise into the cross-correlation S/N estimate \citep[e.g.][]{Wang2017, Hawker2019-lt, Currie2023}. A common practice to reduce the effect of these patterns is to apply a high-pass filter to flatten the off-peak CCF coefficients \citep[e.g.][]{Hawker2019-lt, Currie2023}. However, the filter size that optimally reduces these effects can vary with observation, and may be unknown. In this work, we instead use our knowledge of the model spectrum to predict the shape of the aliasing pattern by cross-correlating our single molecule template spectrum with a noiseless model of the planetary spectrum that includes all molecules. In a similar process to the telluric removal step (Section~\ref{sec:telluric_removal}), this planetary model CCF is then fit and subtracted from the observed data CCF, which leaves a set of residuals, sans aliasing pattern. This process is shown in the middle and lower panels of Figure~\ref{fig:ccf}, where the green CCF is the planetary model CCF, and the black curve shows the detrended residuals. We use the standard deviation of the off-peak detrended residuals as an estimate for noise in the CCF S/N calculation (Section~\ref{sec:SNR}).

\subsubsection{S/N estimation}\label{sec:SNR}
Finally, we measure the S/N of the telluric-free, detrended cross-correlation function by dividing the correlation coefficient at 0 km/s (red region in the lower panel of Figure~\ref{fig:ccf}) by the standard deviation of the detrended residuals (black CCF in the lower panel of Figure~\ref{fig:ccf}). For N hours observed, we integrate N hours of data with random noise, and calculate the cross-correlation detection significance. We then repeat this process for 1000 iterations of noise, enough to ensure the median S/N changes by $<1\%$ with additional iterations. We report the median detection significance for a range of observational scenarios in Section~\ref{sec:results}.

\section{Results}\label{sec:results}
Using our updated SPECTR pipeline, we estimate the detectability of a range of molecular features in spectra of terrestrial planet atmospheres on Earth-sized planets orbiting M dwarf stars, and for a sub-Neptune. We calculate the detectability of molecular features as a function of a grid of relevant parameters, including the observed molecular band, the distance to the system, the performance of the instrumentation, the radial velocity of the star/planet, and the total collecting area of the observatory. Finally, we estimate the accessibility of the most detectable molecular features in a variety of atmospheres for planets in systems analogous to the closest known stellar targets. 
%These targets have a habitable zone planet with a radial velocity minimum mass in the terrestrial range, and  include Proxima Centauri \citep{Anglada-Escude2016-ai}, GJ 1061 \citep{Dreizler2020-ii}, Teegarden's Star \citep{Zechmeister2019-pa}, and GJ 1002 \citep{Suarez_Mascareno2022-hd}. 
We note that all integration/exposure times quoted in our figures and text are defined as the time spent integrating on the target--- the additional time needed for set-up, calibrations, etc. is not included in this estimate. The overhead and real-world time required to obtain hours of observations may well exceed the effective time spent on-sky observing the target.

\subsection{Molecular Detectability}\label{sec:molecular_detectability}
We test the detectability of the major molecular species with spectral features between  0.5--2.0 $\mu m$ (see Table~\ref{tab:bands}) for self-consistent atmospheres of Earth through time, biosignature false positive worlds, and a sub-Neptune atmosphere. For reference, we show planetary flux spectra at the top of the atmosphere for selected molecular bands in Appendix~\ref{apx:spectra}.  Figure~\ref{fig:dot_plot} shows the exposure time in hours required to obtain a $3 \sigma$ detection of each molecular band  for terrestrial atmosphere types orbiting M2V-M8V host stars 1.3 pc away, including both the Earth through time and biosignature false positive scenarios. We note that because the simulated systems in Figure~\ref{fig:dot_plot} are 1.3 pc away and we are assuming $0.5-2\mu$m observations, none of the observed bands fall within the IWA; however, IWA limitations become relevant for observations beyond 1.3 pc (see Sections~\ref{sec:res_nearbytargets},~\ref{sec:iwa_disc}, and~\ref{sec:distance_dep}). The observational configuration assumed for Figure~\ref{fig:dot_plot} includes instrumentation capable of achieving $10^{-3}$ and $10^{-4}$ instrumental (raw) contrast and R=100,000 resolving power, which is reflects the technology goals for the next generation of instruments for the ELTs \citep{Snellen2015-tu} including the GMT \citep{Szentgyorgyi2014-xj}, TMT \citep{Mawet2019}, and ELT \citep{Marconi2022, Palle2023}, and the VLT \citep{Lovis2022}. Figure~\ref{fig:subneptune_detect} similarly shows molecular detectability for the major species in the sub-Neptune atmosphere for a similar observational configuration. We also test the detectability of a range of molecules for analogs of nearby habitable zone targets with atmosphere types including Earth through time, biosignature false positive cases, and a sub-Neptune, and we present the results for the European ELT in Figure~\ref{fig:case_studies}, and for the GMT and TMT in Appendix~\ref{apx:GMGTMTres}. Finally, we test molecular detectability as a function of system RV (Figure~\ref{fig:rv}), and for grids of distance to the system, instrument contrast, and total collecting area in Appendix~\ref{apx:dependencies}.

\begin{figure*}
    \centering
    \includegraphics{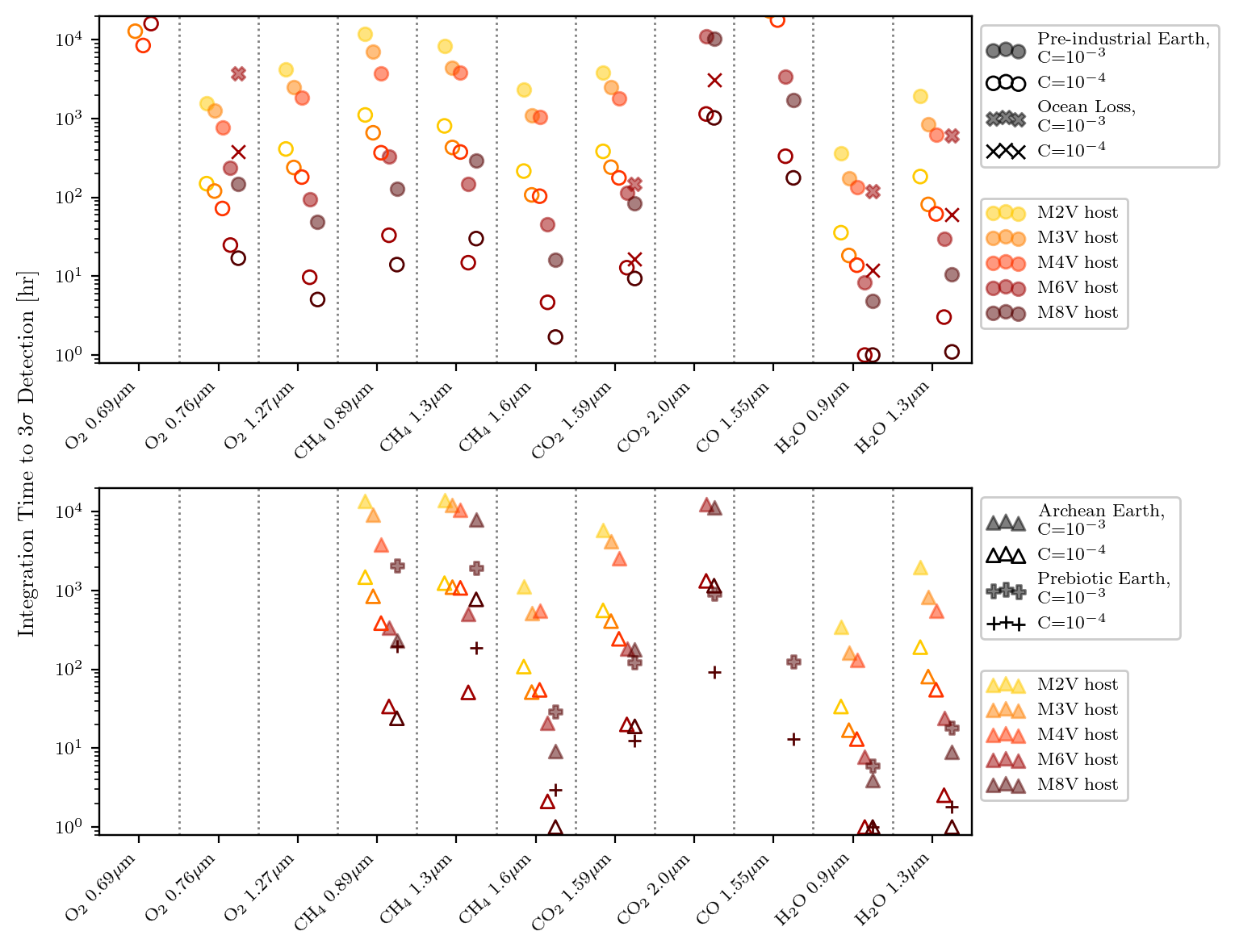}
    \caption{ELT integration time required to obtain a $3 \sigma$ detection of a range of molecular bands for terrestrial atmospheres orbiting hypothetical 1.3 pc distant M2V--M8V host stars. Circular and triangular markers represent pre-industrial and Archean Earth-like atmospheres, respectively. ``X'' and ``+'' markers represent uninhabited ocean loss and prebiotic Earth atmospheres, respectively. Solid and open markers represent assumed instrumental contrasts of C=$10^{-3}$ and C=$10^{-4}$, respectively. We have limited the y-axis to $\sim 10^4$ hr, i.e. $\sim 1$ year of continuous observation. This observational scenario assumes each planet receives 66\% of Earth's insolation, the system has a radial velocity of 20 km/s and is 1.3 pc away from Earth, and the observations are made with the E-ELT with an instrument capable of R=100,000 observations  and total throughput of 10\%. 
    %We emphasize that even though early type host stars are overall more luminous than later type hosts and thus their habitable zones are farther away, planets in their habitable zones receive (and thus reflect) a similar amount of stellar light regardless of spectral type.   
    While the biosignature disequilibrium pairs \ce{O2}/\ce{CH4} and \ce{CO2}/\ce{CH4} are potentially accessible for inhabited worlds, the abiotic gases \ce{O2} and \ce{CH4} in the uninhabited worlds may also be detectable in false positive scenarios.}
    \label{fig:dot_plot}
\end{figure*}

\begin{figure}
    \centering
    \includegraphics[width=0.45\textwidth]{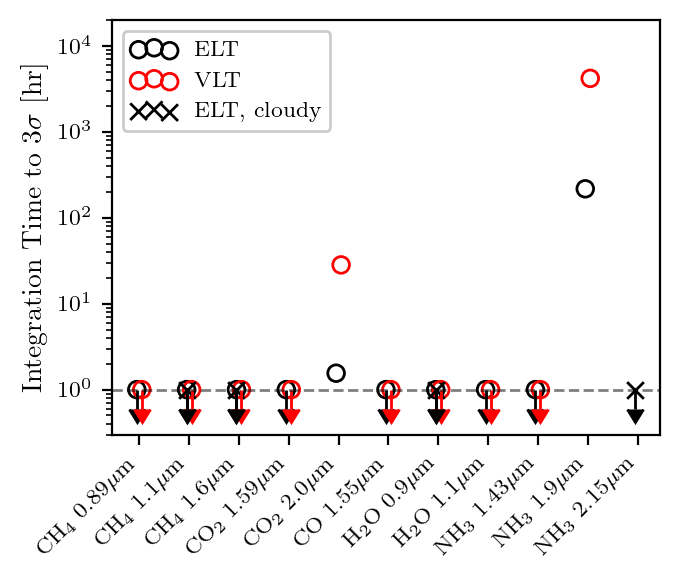}
    \caption{Integration time required to obtain a $3 \sigma$ detection of a range of molecular bands for a sub-Neptune atmosphere \citep{Charnay2015} orbiting Proxima Centauri b using simulated E-ELT and VLT observations with an assumed achieved contrast of C=$10^{-4}$ and total throughput of 10\%. The molecules in this atmosphere are highly detectable with the ELT, requiring less than a few hours for most molecular species. The overall high detectability of all molecules, including H-bearing species, %may help break the Msin(i) degeneracy 
    may help to identify the atmosphere of Proxima Centauri b as more likely to be sub-Neptune-like, rather than terrestrial, and this may be possible in the near term with smaller aperture observatories. Note: Markers on the horizontal line at 1 hr (with arrows) are upper limits. }
    \label{fig:subneptune_detect}
\end{figure}

\subsection{Inhabited Worlds and Biosignatures}
\begin{figure}
    \centering
    \includegraphics{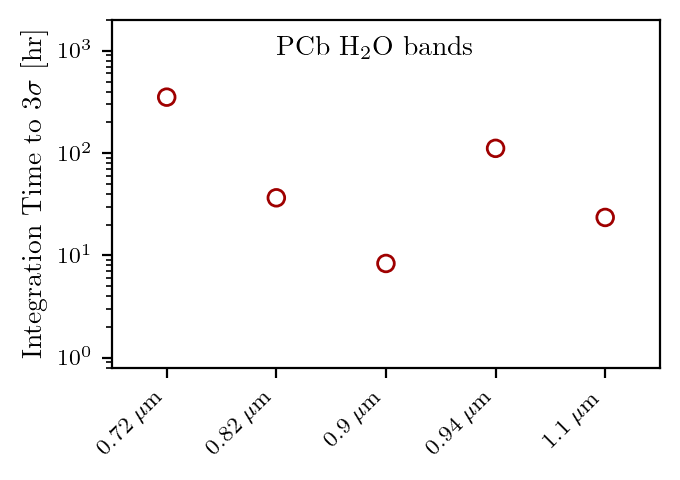}
    \caption{Detectability of \ce{H2O} bands spanning the NIR for Proxima Centauri b with a modern Earth-like atmosphere. We assume an instrumental contrast of $10^{-3}$ and R=100,000.  We find that the 0.9 $\mu$m band is the most detectable \ce{H2O} feature by at least a factor of three in some cases, and by 1--2 orders of magnitude in the worst cases. }
    \label{fig:h2o_bands}
\end{figure}

Examining the detectability of molecular bands for different, yet equidistant, stellar hosts reveals that molecular features are more detectable around later type stars overall, and that the most sensitive band for a given molecule can change with host star spectral type. Figure~\ref{fig:dot_plot} presents the detectability of selected molecular features for planets in systems at 1.3pc, for the inhabited pre-industrial Earth-like planet (circles) and its ocean loss false positives for an M6V planet (``X'' markers) (upper panel) and for the Archean Earth-like planet (triangles) and its M8V prebiotic high-volcanism false positive (``$+$'' markers)  (lower panel) atmospheres, respectively. Except for the 0.69$\mu$m \ce{O2} band, all the molecular features are easier to detect for later type stars, due to the more favorable contrast between planet and star.  For the pre-industrial Earth-like atmospheres, \ce{O2}, \ce{CO2}, \ce{CH4} are all detectable in $<100$ hours for late-type hosts, and between 100--1000 hours for early type hosts for the most optimistic instrumental contrast $C=10^{-4}$ cases, and may reqiure up to an order of magnitude more time for $C=10^{-3}$ cases. For the M6V host at 1.3 pc (analogous to Proxima Centauri b), our results suggest that the most detectable bands of \ce{O2}, \ce{CO2}, \ce{CH4}, and \ce{H2O} are the 1.27, 1.59, 1.6, and 0.9 $\mu$m bands, respectively. At the minimum, \ce{O2}, \ce{CO2}, \ce{CH4}, and \ce{H2O} will require approximately 10, 13, 5, and 1 hours to detect, respectively, with $C=10^{-4}$.  \ce{H2O} at 0.9 $\mu$m is the most accessible molecular feature for pre-industrial Earth-like cases in our work--- we additionally test the detectability of shorter wavelength \ce{H2O} bands, including the 0.72 and 0.82 $\mu$m bands, which, in addition to being more likely to  be outside the IWA, are not as affected by the telluric absorption and may benefit from a stronger continuum flux due to Rayleigh scattering; however, we still find the 0.9 $\mu$m band to be the most detectable feature (see Figure~\ref{fig:h2o_bands}). We note that while the 0.82 $\mu$m \ce{H2O} feature requires roughly 3x more observation time to detect, it is more likely to fall outside of the IWA due to its location at shorter wavelengths. In cases with early-type hosts, the \ce{O2} A-band at 0.76 $\mu$m is more accessible than the 1.27 $\mu$m NIR band because the SED of the host star peaks at shorter wavelengths, and there are more reflected photons in the 0.76 $\mu$m range than for the 1.27$\mu$m range. Similarly, for late-type hosts the NIR band may be the more detectable target because the  SED of the host star peaks at longer wavelengths, and there are more NIR photons than visible.  A similar effect was highlighted for transiting planets in \citet{Currie2023}.  \ce{CO} may be challenging due to its low abundance ($<1$ ppm) in this atmosphere type, requiring $>10^3$ hours to detect. \ce{O3} does not have the high-resolution structure required for detection via cross-correlation in the wavelengths we tested, as discussed in \citet{Currie2023}.

The molecular features in the Archean Earth atmospheres we tested have similar detectablity when compared to the pre-industrial Earth cases, with the exception of \ce{O2} which is at sufficiently low abundance that it is not detectable in the Archean atmosphere, and \ce{CH4}, whose higher abundance is more detectable. The lower panel of Figure~\ref{fig:dot_plot} (triangular markers) shows that %\ce{CO2}, \ce{CH4}, and \ce{H2O} are all accessible for Archean worlds.  
for an Archean Proxima Centauri b, \ce{CO2} and \ce{CH4} are detectable in approximately 20 and 2 hours, respectively, and \ce{H2O} is accessible in an hour of observation for an optimistic instrumental contrast of $C=10^{-4}$, and up to an order of magnitude more time for (perhaps more realistic for first-light instrumentation) $C=10^{-3}$ cases. CO is challenging due to a low abundance, and the Archean atmospheres have little to no \ce{O2} and \ce{O3} available to detect. 

\subsection{Uninhabited Worlds and False Positive Discriminants}
In addition to the inhabited planet results, the upper and lower panels of Figure~\ref{fig:dot_plot} also show detectability results for two biosignature false positive scenarios, including an ocean loss world \citep{Luger2015, Meadows2018-yx, Leung2020} (upper panel, ``X'' markers) and a prebiotic Earth atmosphere \citep{Krissansen-Totton2018-tj, Wogan2020, Meadows2023} (lower panel, ``$+$'' markers). These biosignature false positive scenarios can build up potentially detectable levels of abiotically generated \ce{O2} in the case of the ocean loss world, and \ce{CO2} and \ce{CH4} in the prebiotic Earth scenario. Our results suggest that for the ocean loss scenario, \ce{CO2} is accessible in $\sim 17$ hours at a contrast of C=$10^{-4}$, similar to the pre-industrial Earth case because of the strong absorption of the \ce{CO2} band. While the pre-industrial Earth around an M6V star had an \ce{O2} A-band at 0.76 $\mu$m that was less detectable than the 1.27 $\mu$m \ce{O2}-band, its ocean-loss  false positive has an \ce{O2} A-band at 0.76 $\mu$m takes approximately two orders of magnitude less time than the 1.27 $\mu$m band (380 and 60,000 hours for the 0.76 and 1.27 $\mu$m bands, respectively for the $C=10^{-4}$ cases).  Our results confirm those of \citet{Leung2020}, who also modeled massive 10-bar \ce{O2}-dominated post-ocean-loss atmospheres, and showed that the 1.27 $\mu$m \ce{O2} band will be suppressed due to \ce{O2}--\ce{O2} collisionally-induced absorption. However, we note that all \ce{O2} bands, except for the weaker 0.69 $\mu$m band, will be more  challenging to detect than for the pre-industrial Earth case.  Water vapor is also more challenging to detect for the ocean loss world, despite comparable surface water abundances, requiring 11 hrs for water detection compared to the pre-industrial Earth-like case's 1 hr for optimistic $C=10^{-4}$ cases.  
%since this world is undergoing ocean loss (but still retains an ocean) and the atmospheric \ce{H2O} is being photolyzed, 
This is due primarily to \ce{H2O} line broadening in the dense \ce{O2} atmosphere, which reduces the effective number of absorption lines accessible to the cross-correlation, as well as some suppression of the 0.94 $\mu$m \ce{H2O} band due to \ce{O2}--\ce{O2} collisionally-induced absorption. This lack of sensitivity to broad absorption features is a consequence of the cross-correlation technique, which requires many resolved absorption lines to be effective; thus, these broad absorption features may be detectable using a different signal extraction technique at lower resolution.
In summary, our results suggest the ``false positive'' \ce{O2} in the ocean loss case may be more challenging to detect than the biological \ce{O2} in the pre-industrial Earth case, with the characteristic that the 0.76 $\mu$m band is easier to detect than the 1.27 $\mu$m band, and \ce{H2O} may not be as readily detectable for this \ce{O2} false positive scenario. 

%% Prebiotic world: CO2 and CH4 are accessible, O2 is not but that was expected. The discriminant here would be CO, but that might be challenging as it will take ~1000 nights for the nearest star (Fig 11). This is a legit false positive we might have to worry about... see more in discussion. 
The prebiotic Earth scenario (lower panel, ``$+$'' markers in Figure~\ref{fig:dot_plot}), can generate detectable \ce{CH4} and \ce{CO2} in its atmosphere, which can be a false positive for the \ce{CH4}/\ce{CO2} biosignature pair \citep{Krissansen-Totton2018-tj, Meadows2023}. Both \ce{CH4} and \ce{CO2} are similarly detectable near 1.6 $\mu$m in the prebiotic Earth case as in the Archean Earth case, requiring approximately 3 and 12 hours of observation, respectively,  %Therefore, the origin of \ce{CH4} must be constrained to use this pair of molecules as a biosignature.
and \ce{H2O} is similarly detectable in both cases and  accessible within a few hours of observing for the optimistic instrumental contrast $C=10^{-4}$ cases, and these estimates increase by up to an order of magnitude for $C=10^{-3}$ cases. The discriminant between the two cases is the significantly more detectable  \ce{CO} in the prebiotic Earth case, which requires only 13 hours of observing time at 1.55 $\mu$m, as opposed to $\sim10^4$ hours in the inhabited Archean case for $C=10^{-4}$ cases. This high abundance of CO relative to \ce{CO2} and \ce{CH4} is an indicator that the \ce{CH4} could be volcanic \citep{Wogan2020}. Additionally, we note that in the cases of high CO abundance, the 1.55 $\mu$m band is more detectable than the 2.3 $\mu$m band, despite the 2.3 $\mu$m band being the stronger absorber of the two.  Observing the \ce{CO} band at $2.3 \mu$m has several additional challenges, including an increased sky background brightness, a decreased stellar SED continuum, and IWA limitations  (see Section~\ref{sec:iwa_disc}) that combined make detecting the 2.3 $\mu$m \ce{CO} band more challenging than the 1.55 $\mu$m band. We discuss the implications of molecular detections and non-detections for both the inhabited and uninhabited worlds in Section~\ref{sec:discussion}.

\subsection{Sub-Neptune type atmosphere}\label{sec:subnepresults}

For our simulations of Proxima Centauri b as a sub-Neptune, we find that both the E-ELT and the VLT need of the order of an  hour to make a 3-$\sigma$ detection of at least one band of several molecules (Figure~\ref{fig:subneptune_detect}), including \ce{NH3}, which can be diagnostic of an H-dominated atmosphere. 
In particular, multiple bands of \ce{CH4}, \ce{CO2}, \ce{CO}, and  \ce{H2O}, as well as and the 1.43 $\mu$m band of \ce{NH3},  may be detectable in as little as an hour of observation time using the E-ELT or VLT. We tested the detectablity of these molecular features for both the cloudy and clear cases simulated by \citet{Charnay2015}, and find that the detectability of most of the molecules in the cloudy atmosphere are within a few percent--- cloudiness does not significantly impact molecular detectability in most cases. This is because high-resolution spectroscopy is sensitive to line cores that allow us to probe higher altitudes above the cloud deck, an improvement over low-resolution spectroscopy where the clouds can raise the spectral continuum and obscure spectral features \citep{Gandhi2020}. The high detectability of species in the  GJ 1214 b atmosphere is likely due to two main factors: 1) GJ 1214 b's roughly 3x larger diameter than Earth, which increases the flux of reflected photons, 2) the higher abundances of H-bearing species in the sub-Neptune atmosphere (e.g. $\sim10^4$ ppm \ce{CH4}, vs. $\sim 100$ ppm \ce{CH4} and little to no \ce{NH3} in the Archean atmosphere) \citep{Charnay2015}.

\subsection{Prospects for Simulated Nearby Targets}\label{sec:res_nearbytargets}

To understand the potential impact of target distance on our detection sensitivity and usable wavelength range, we simulate observations of hypothetical nearby habitable zone targets orbiting a mid-type M dwarf (M6V) across a grid of distances between 1.3 pc and 5 pc (Figure 8). This grid of cases roughly encompasses several known real targets that may host $\sim$Earth-sized planets in or near the habitable zone, including Proxima Centauri b \citep[1.3 pc distant, M6V host][]{Anglada-Escude2016-ai}, GJ 1061 d \citep[3.67 pc, M5.5V host][]{Dreizler2020-ii}, Teegarden’s Star c \citep[3.83 pc, M7V host][]{Zechmeister2019-pa}, and GJ 1002 b and c \citep[4.85 pc, M5.5V host][]{Suarez_Mascareno2022-hd}. However, we note that the biosignature and false positive atmospheres simulated here are for a planet placed in the host star's habitable zone at 67\% of Earth's insolation. This placement  does not necessarily match the orbital distance of the known planets orbiting these stars, although this is a good approximation for Proxima Centauri b \citep{Anglada-Escude2016-ai}. For the other targets, this approximation may lead to under or overestimates of abundances of photochemically and climatically controlled gases such as \ce{CH4}, \ce{O3}, and \ce{H2O} \citep{Grenfell2007}. We present this grid of cases in Figure 8, where we show the integration time to a 3$\sigma$ detection for all atmosphere scenarios we consider in this work, and for instrumental contrasts of C=$10^{-3}$ (filled markers) and $10^{-4}$ (open markers). We consider an assumed IWA of 2 $\lambda/D$, and do not plot molecules that are at wavelengths within the IWA for a given planet. We limit the y-axis to $\sim10^4$ hr (i.e. one year of continuous observation), and thus missing markers indicate cases where the distance to the system hinders our ability to detect the molecular band either via IWA limitations or low target flux due to the distance inverse square law; additionally, missing markers in some cases may indicate that the molecule is not present in the atmosphere, e.g. \ce{O2} in the Archean Earth-like cases, where no markers for any distance will be plotted for these molecules.

We find that our Proxima Centauri b analog at 1.3 pc away is likely the best target for atmospheric characterization using the ELTs, and our ability to characterize habitable zone targets farther away than this may be significantly hindered by IWA limitations. With the E-ELT, it may be possible to detect \ce{H2O} in as little as an hour for a terrestrial atmosphere on the planet Proxima Centauri b with an instrumental contrast of $C=10^{-4}$, and in  $\sim 10$hr for $C=10^{-3}$ (Figure 8). This  is comparable to the time required to detect \ce{H2O} and many other species for a sub-Neptune atmosphere (Figure 6). Other molecular species in terrestrial atmospheres at the distance of PCb such as \ce{O2}, \ce{CO2}, \ce{CH4}, and \ce{H2O} may be detectable on the order of $\sim 10-100$ hours of observing, depending on the contrast achieved.

The detectability of molecular spectral features for mid-type M dwarf targets beyond 1.3 pc away may be significantly affected by the distance/IWA limitation, and thoroughly characterizing these targets may be challenging, but planetary targets around earlier type M dwarfs may pose less of a challenge. Our ability to probe the NIR molecular bands deteriorates with distance primarily due to the IWA constraints, and detecting the \ce{CO2}, \ce{CH4}, and \ce{CO} bands may become challenging beyond $\sim 3$pc for $\sim$M6V systems.  However, the \ce{O2} A-band, the \ce{H2O} 0.9 $\mu$m band, and \ce{CH4} bands between 0.8–0.9 $\mu$m may still be detectable for targets out to $\sim3--4$pc with significant observational resources, requiring 10--100s of hours for \ce{H2O}, and 100--1000s of hours for \ce{O2} and \ce{CH4}, depending on achieved instrumental contrast.  This would allow us to search for some signs of habitability and life on these more distant targets, but gaining the full environmental context for these systems (the GJ 1061, Teegarden's Star, and GJ 1002 systems, for example)  may be more challenging than for a 1.3 pc distant case such as Proxima Centauri b. We discuss the effect of distance further in Appendix~\ref{apx:dependencies}. We note, however, that earlier type targets may be less hindered by the IWA limitation , and may also benefit from improved AO contrast at larger separations. 

\begin{figure*}
    \centering
    \includegraphics{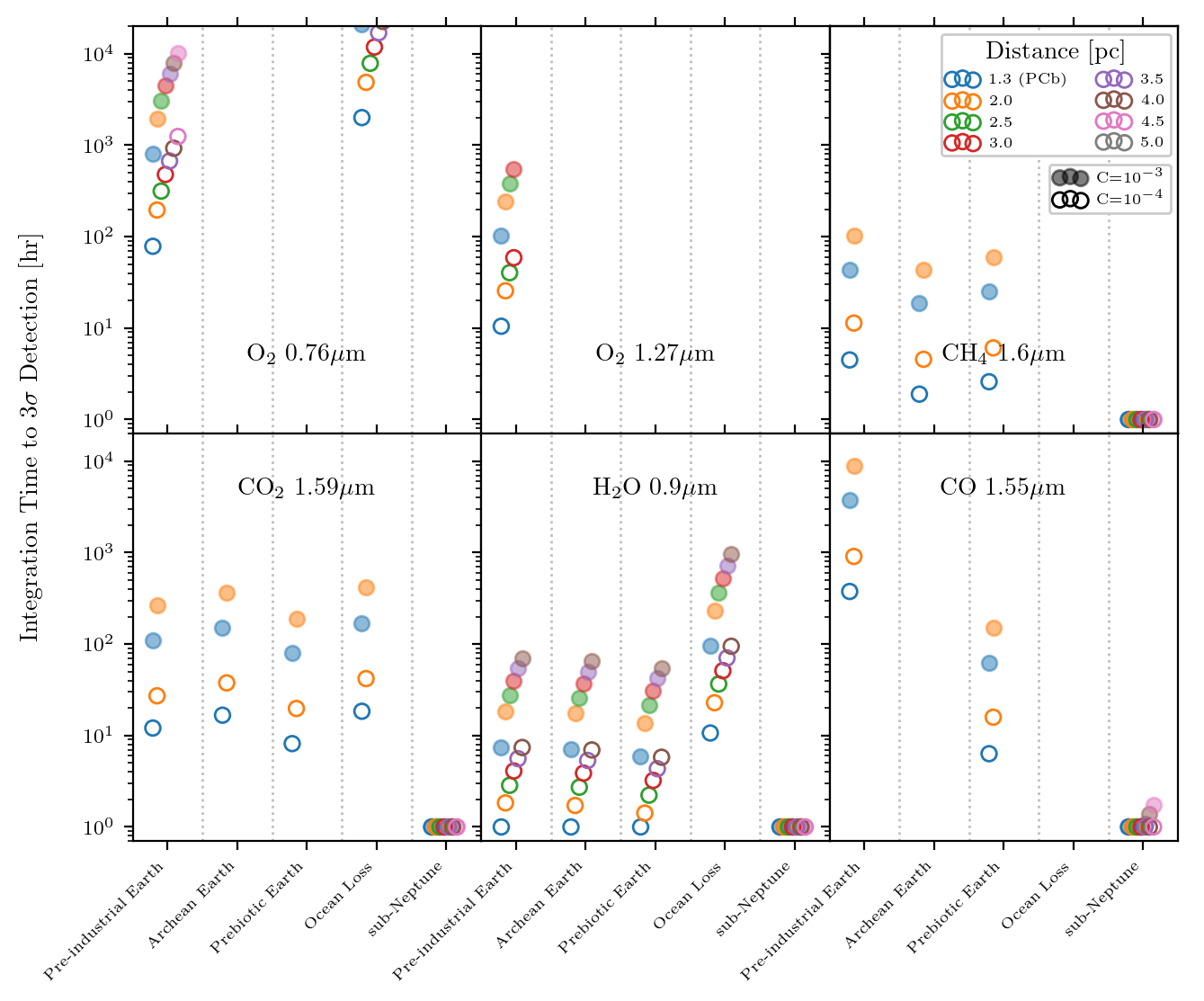}
    \caption{Integration time to a $3 \sigma$ detection for a range of terrestrial atmosphere types on Earth-sized planets orbiting analog nearby targets. Each panel represents a different molecular band, and each color represents a different nearby host star/planet analog. Filled and open circles represent assumed instrumental contrasts of C=$10^{-3}$ and C=$10^{-4}$, respectively, and we assume a total throughput of 10\%. A Proxima Centauri b analog (1.3 pc away) is plotted in blue, and a sub-Neptune version of Proxima Centauri b is included for comparison.  We may be able to detect several molecules within 100 hours of observing each target, and our Proxima Centauri b analog is the most favorable nearby target for thorough atmospheric characterization.}\label{fig:case_studies}
\end{figure*}

\subsection{System RV Dependence}\label{sec:rv}
\begin{figure}
    \centering
    \includegraphics{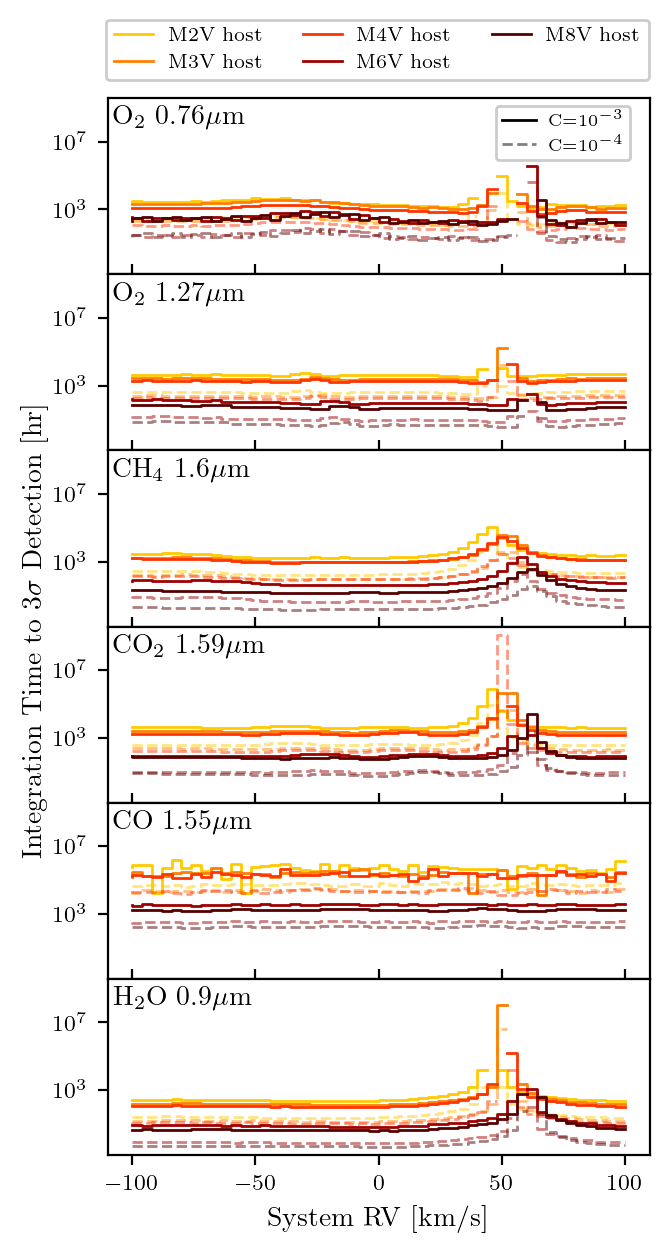}
    \caption{Time to a $3 \sigma$ detection as a function of system RV for pre-industrial Earth atmospheres orbiting M dwarf hosts.  The host star is assumed to be 1.3 pc away, and observed with E-ELT instrumentation capable of R=100,000. Solid and dashed lines represent assumed instrumental contrasts of C=$10^{-3}$ and C=$10^{-4}$, respectively. We assume zero barycentric velocity and a total throughput of 10\%. Our ability to detect molecular features depends strongly on the degree of telluric/target line blending, and will be a consideration for minimizing the total observation time to detect molecular features in future observing campaigns by either vetting targets for non-optimal systemic velocities, observing the planet at orbital positions that minimize line blending, or taking into account observational seasonality if the barycentric velocity plays a significant role for a particular target. }
    \label{fig:rv}
\end{figure}

Following \citet{Lopez-Morales2019} and \citet{Hardegree-Ullman2023}, we explore the impact of system RV on integration time, assuming zero barycentric velocity. Figure~\ref{fig:rv} shows the effective observation time to a $3 \sigma$ detection as a function of system RV for pre-industrial Earth-like planets around M2V--M8V host stars at approaching quadrature.  This simulation assumes that the target system is at a distance of 1.3 pc, and is observed with the E-ELT with instrumentation capable of R=100,000 and C=$10^{-4}$ (dashed lines) or C=$10^{-3}$ (solid lines). The total RV of the planet can play a large role in the detectability of a molecular feature because it controls the degree of separation or overlap between the target and telluric features, which can prevent planet/telluric line separation \citep{Lopez-Morales2019, Hardegree-Ullman2023}. In practice, the instantaneous RV of the planetary target will depend not only on the intrinsic radial velocity of the host system, but also the barycentric radial velocity, i.e. the radial velocity due to the Earth's position around the Sun relative to the target, and the RV of the planet as a function of its orbital position. The planet's instantaneous radial velocity can be on the order of the host star's radial velocity, especially at quadrature, and our ability to separate the telluric and target spectra will depend strongly on both the host star and planetary RVs. 

While the planetary RV changes as it orbits its host star, the host star RV (systemic RV), is fixed. To illustrate the dependency on the system RV of the host star, we hold the planet at approaching quadrature where its maximum RV occurs and test the molecular detectability over a grid of host star system RVs in Figure~\ref{fig:rv}.  In most panels of Figure \ref{fig:rv}, a spike in observation time occurs near $\sim 50$km/s and spans $\sim 10$ km/s, due to significant line blending between the telluric and target spectra that significantly hinders detection, a result also seen by \citet{Lopez-Morales2019} and \citet{Hardegree-Ullman2023}. The spikes in observation time for the different host stars are slightly offset from one another because the planet's orbital distance, and thus its RV at quadrature, is set according to the position of the habitable zone--- the habitable zones of late type stars are closer in than early type stars. Note that the panel in Figure~\ref{fig:rv} showing CO detectability does not show maxima near 50 km/s as in the other panels---any line blending that occurs due to the systemic RV does not significantly affect its detectability, because of the low telluric CO abundance ($\sim 50-100$ ppb). In summary, as in \citet{Lopez-Morales2019} and \citet{Hardegree-Ullman2023}, we find that total observation time may be reduced by minimizing the level of line blending for a given target. This can be achieved by using a higher spectral resolution instrument, selecting for targets with optimal systemic velocities, observing the planet at different orbital positions that minimize the amount of line blending, or observing the target seasonally if the barycentric RV has a significant Doppler shifting effect on the target; however, we note that for our simulations we assume zero barycentric velocity.

\subsection{Spectral Resolution}
While we do not explicitly model the effect of increasing the spectral resolution of our observations (we assume a fixed resolution of $R=100,000$ for the observations we consider in this work), we discuss qualitatively how spectral resolution could impact reflected light spectroscopy observations. Varying the spectral resolution for transit transmission spectra has been investigated extensively by \citet{Rodler2014} and \citet{Lopez-Morales2019}, and later revisited by \citet{Hardegree-Ullman2023} and \citet{Currie2023}, who all find that increasing the spectral resolution of the observation can provide several distinct advantages for transiting planets. \citet{Lopez-Morales2019} report that increasing the spectral resolution of the observation from $R=100,000$ to $R=500,000$ can reduce the level of blending between the telluric and planetary spectral lines by up to a few times in both amplitude and width of the blended regions due to the increased number of spectral lines resolved, relaxing the RV restrictions discussed above. This can also more than double the depths of the spectral lines, which can reduce the total time to achieve a detection by over 30\%. The spectral resolution we can achieve in practice, however, will be set by the detector dark current, which will be especially relevant for faint rocky exoplanet targets. We note that these studies \citep{Lopez-Morales2019, Currie2023, Hardegree-Ullman2023} only test transit transmission spectra, which is sensitive to the upper atmosphere where absorption lines are narrow. For reflected light spectra, which can probe deeper into the atmosphere, the absorption lines will be broader due to higher pressures and temperatures. This may reduce the advantages for increasing spectral resolution beyond R$\sim100k--150k$, which regardless may be difficult to achieve without losing throughput and wavelength coverage. Increasing spectral resolution does not necessarily provide the same advantages for reflected light spectra, and for this work we merely point out this complexity qualitatively, leaving a more thorough investigation for a future study.

% \begin{deluxetable*}{c|ccccc}\label{tab:realtargets}
% \tablecaption{Nearby habitable zone planet system properties}
% \tablewidth{0pt}
% \tablehead{
% \colhead{Target Analog} & Distance& System RV   & Analog Stellar Host Type & Planetary Radius & Separation \\
%  & [pc] & [km/s] &  & [M$_\oplus$] & [mas] 
% }
% \startdata
% Proxima b & 1.3$^a$ & -22$^b$ & M6V & 1 & 31 \\
% GJ 1061 d & 3.67$^c$ & 1$^d$ &M6V & 1 & 11 \\ 
% Teegarden's Star c & 3.83$^e$ & 68$^f$ & M4V & 1 &  42 \\
% GJ 1002 b/c & 4.85$^g$ & -40$^h$ & M6V  & 1 & 8.3\\
% \enddata
% \tablerefs{$^a$\citet{Anglada-Escude2016-ai}, 
%             $^b$\citet{Kervella2017-xo},
%             $^c$\citet{Dreizler2020-ii},
%             $^d$\citet{Vallenari2023-op},
%             $^e$\citet{Zechmeister2019-pa},
%             $^f$\citet{Tanner2012-hu},
%             $^g$\citet{Suarez_Mascareno2022-hd},
%             $^h$\citet{Vallenari2023-op},
%             }
% \end{deluxetable*}

\section{Discussion}\label{sec:discussion}
Our results show that a range of molecular features that can be used to identify terrestrial planets, and that serve as signs of  habitability, biosignatures and abiotic processes,  may be detectable in ELT reflected light observations of nearby exoplanets.  Here we discuss what could be learned about terrestrial exoplanets from these observations, as well as recommendations for optimal targets and observing considerations and protocols.  

\subsection{Characterizing Terrestrial Exoplanets in Reflected Light From the Ground}

The ELTs will provide the first opportunity to identify and study the atmospheres of non-transiting terrestrial exoplanets, thereby enabling the search for atmospheres, habitability and life.  These near-term observations will yield valuable lessons learned for the development of future instrumentation and the upcoming Habitable Worlds Observatory NASA flagship mission.

\subsubsection{Discriminating Terrestrials and Sub-Neptunes}\label{sec:subneptunes}

Determining whether a planet is in fact terrestrial and not a sub-Neptune is the first step in terrestrial exoplanet characterization, and this process will not necessarily be straightforward for direct imaging targets, whose size and true mass may be unknown. Transit observations are potentially less ambiguous, as they can provide precise measurements of a planet's size \citep[e.g.][]{Seager2003}, and statistical arguments based on the known properties of similar sized planets can then be used to infer whether or not a planet is terrestrial \citep[e.g.][]{Rogers2015, Martinez2021}.  However, for direct imaging targets in reflected light,  the size of the planet will be degenerate with its albedo, and although extremely large targets can likely be ruled out, planetary brightness alone will likely lack the precision needed to discriminate between terrestrial and sub-Neptune sized planets for directly imaged targets \citep[e.g.][]{Feng2018, Carrion2020}.  Similarly many non-transiting, direct imaging targets are discovered with radial velocity, which can only return a mass that is a lower limit, due to the unknown inclination and therefore observed radial velocity component of the planet's orbit \citep{Hatzes2016}. %And while thermal or reflected light phase curves may reveal a planet's inclination and true mass, thermal measurements of a habitable terrestrial planet are unlikely to be possible with JWST in the near term, given the relatively small day-night contrast of a habitable zone planet with an atmosphere \citep{Kreidberg2016}, and  JWST's lack of sensitivity to temperate planets in the habitable zones of their parent stars \citep{Meadows2018-gh,Lustig-Yaeger2019-bk}. 
It has been proposed that determining the planetary mass and inclination through high-resolution spectral observations could be achieved by either monitoring the planetary orbit via molecular detections over a timeseries of observations and fitting an orbital model \citep[e.g.][]{Rodler2012, Hoeijmakers2018betapic}, or observing the planet at quadrature to measure the planetary semi-amplitude and using the RV-measured stellar semi-amplitude to solve for the planetary mass and inclination \citep{Lovis2017-rr}. Of course, this method of determining the orbit with high-resolution spectroscopy may only be necessary if the planetary spectrum is significantly embedded in the stellar speckle noise; in a more standard direct imaging scenario, monitoring the orbital motion of the planet will likely be enough to fit a model to determine the orbital parameters.   However, a measured mass, or mass range, that spans the two planetary populations may leave these measurements ambiguous as to the nature of the planet, including its radius and/or atmospheric composition.  Here, we instead investigate discriminating terrestrials from sub-Neptunes by characterizing the atmosphere via spectroscopy to more definitively confirm or rule out a sub-Neptune. %could be definitively performed using spectroscopy confirm or rule out a sub-Neptune.
 
Our results show that sub-Neptunes may be readily discriminated from potentially terrestrial planets with high-resolution direct imaging spectroscopic observations of reduced gases, and that the ambiguous nature of RV-detected Proxima Centauri b could be conclusively resolved in as little as a night of observing.  We also show that \ce{CO2} and \ce{H2O} are comparatively detectable in sub-Neptune and terrestrial atmospheres, and so are poor discriminants for terrestrial worlds.  For the sub-Neptune atmosphere used in this study, which is shown in Figure \ref{fig:spaghetti_subneptune} \citep{Charnay2015}, our results show that the molecular bands of reduced gases including \ce{NH3} and \ce{CH4}, and CO, may be detectable at a 3-$\sigma$ level in as little as an hour of observation for bands between 0.7 and 1.4 $\mu$m (Figure~\ref{fig:subneptune_detect}).  Moreover, reduced gases such as \ce{NH3} are much less likely to be seen in an oxidizing terrestrial atmosphere, such as the Earth's atmosphere throughout time, and the detection of reduced gases may allow us to discriminate an atmosphere with a primarily reducing atmosphere from one with a primarily oxidizing atmosphere. We also found that \ce{CO2} was more detectable in the sub-Neptune atmosphere, as its lower abundance was more than compensated by the size of the sub-Neptune.  Water vapor was also comparably detectable for both the sub-Neptune and the 1.3 pc distant terrestrials, requiring as little as an hour of observing time (Figure~\ref{fig:case_studies}), or significantly more detectable in the sub-Neptune than in more distant terrestrials. Searching for molecules indicating a sub-Neptune atmosphere could also be done in the near term using a VLT-sized observatory also for relatively little observing time (Figure~\ref{fig:subneptune_detect}), and ruling out a sub-Neptune atmosphere before the ELTs come online may be a strong observing strategy (see Section~\ref{sec:observingprotocol}). Since abundant water vapor may be a common feature of most sub-Neptune type atmospheres \citep[e.g.][]{Bean2021}, our results suggest that \ce{H2O} and \ce{CO2} alone are poor discriminants for terrestrial vs sub-Neptune exoplanet identification in reflected light observations.  Given that the inclination ambiguity in the RV mass means that there may be at least a 10\% likelihood that Proxima Centauri b is a sub-Neptune \citep{Bixel2017}, the high detectability of the sub-Neptune atmosphere type should allow us to conclusively rule out the sub-Neptune nature of Proxima b with only a few hours of observing. This approach could also be considered for other targets with ambiguous masses.

%Both terrestrial and sub-Neptune type atmospheres may include detectable water vapor (Figure~\ref{fig:case_studies}), and it may be possible to discriminate between the two by detecting reduced gas species in the sub-Neptune type atmosphere (Section~\ref{sec:subnepresults}). . Several molecular features, including those of \ce{H2O}, \ce{CO2}, and \ce{CH4} may be accessible in both sub-Neptune and terrestrial atmospheres at 1.3 pc (Figure~\ref{fig:case_studies}); however, terrestrial atmosphere features may require as much as two orders of magnitude more observation time than sub-Neptune atmospheres. Additionally, it is likely that the reduced gases \ce{NH3} and \ce{H2} are also highly accessible in a sub-Neptune atmosphere (Table~\ref{tab:subneptunes}). \ce{NH3} and \ce{H2} are not expected in an oxidizing terrestrial atmosphere, and the detection of reduced gases may allow us to discriminate an atmosphere with a primarily reducing atmosphere from one with a primarily oxidizing atmosphere. In summary, the high detectability and presence of H-bearing species may allow us to discriminate a sub-Neptune atmosphere from a terrestrial atmosphere in as little as a single night of observing for nearby non-transiting exoplanets. Given that there may be over a 10\% probability that Proxima Centauri b is a sub-Neptune \citep{Bixel2017}, the high detectability of this atmosphere type should allow us to test this with relatively few observational resources.

\subsubsection{Habitability assessment}
For a terrestrial exoplanet the presence of greenhouse gases can provide  insight into its climate and habitability, and we find that the greenhouse gases \ce{H2O}, \ce{CO2} and \ce{CH4} are likely accessible to the ELTs for nearby non-transiting targets. Sufficient atmospheric water vapor can contribute to the greenhouse warming needed to maintain liquid surface water \citep{Kasting1993-cn}, and our results show that 0.9 $\mu$m \ce{H2O} is the most accessible band for most terrestrial targets (Figure~\ref{fig:case_studies}), needing only 1 hr of integration time for a 3-$\sigma$ detection on a terrestrial Proxima Centauri b in the most optimistic case.  Additionally, an \ce{H2O} detection, while not guaranteeing surface liquid water, could make surface liquid water more likely, and liquid water is thought to be critical for developing and supporting life \citep[e.g.][]{Ball2008-ub, Pohorille2012}. Given its relative ease of detection, we recommend the 0.9 $\mu$m \ce{H2O} band as a prime target for initial terrestrial atmosphere characterization efforts using the ELTs. However, as described in the previous section, considering \ce{H2O} as a habitability indicator necessitates a degree of caution until the terrestrial/sub-Neptune degeneracy can be resolved.

\ce{CO2} and \ce{CH4} are also powerful greenhouse gases that both impact planetary climate and support biosignature detection, and these are likely accessible to the ELTs for approximately 3--10x more observing time than \ce{H2O} (on the order 10 of hours for Proxima Centauri b).   In particular, in Earth-like environments \ce{CO2}  may be strongly tied to geological processes and so may also be indicative of an active carbonate--silicate cycle \citep{Walker1981-wj, Kasting1993-cn, Kopparapu2013-wm}, which could buffer the planetary climate through geologic time. As is the case for water though, \ce{CO2} may also be readily detectable in sub-Neptune atmospheres (Figure \ref{fig:case_studies}), and so it is important to discriminate terrestrial from sub-Neptune atmospheres (see Section~\ref{sec:subneptunes}) before attempting to interpret a \ce{CO2} detection. 

\subsubsection{Biosignatures}
We find that two biosignature disequilibrium pairs, \ce{O2}/\ce{CH4} and \ce{CO2}/\ce{CH4}, may be accessible to the ELTs.  Detecting both \ce{O2} and \ce{CH4} would reveal the canonical \ce{O2}--\ce{CH4} biosignature pair, indicating a chemical disequilibrium and evidence for active fluxes from both oxygenic photosynthetic, and methanogenic organisms \citep{Hitchcock1967-ky, Meadows2018-gh}. The presence of both \ce{CO2} and \ce{CH4} may reveal an active flux of \ce{CH4} that is likely higher than expected for Earth-like geological activity alone \citep{Krissansen-Totton2016-bq,Krissansen-Totton2018-tj}.  Our reflected light results show that the biologically-mediated  \ce{CO2}/\ce{CH4} disequilibrium is comparably detectable in both the Archean and PIE atmospheres (within a factor of 2--3) making it a long-lived biosignature, a similar result to that of \citet{Meadows2023} for simulations of JWST transmission observations of TRAPPIST-1.   The \ce{O2}/\ce{CH4} disequilibrium is only detectable in the PIE case, which has a photosynthetic biosphere. \ce{O2}, \ce{CH4}, and \ce{CO2} may require more observation time than \ce{H2O}, but may still accessible within approximately 100 hours for nearby targets with a pre-industrial Earth-like atmosphere. For Proxima b, $\sim 10$ hours of observing would be needed to detect these biosignature pairs if the molecules can be observed simultaneously. The most accessible spectral features for these molecules will be \ce{O2} at both 0.76 $\mu$m and 1.27 $\mu$m, \ce{CH4} at 0.89, 1.3, and 1.6 $\mu$m, and \ce{CO2} at 1.59 $\mu$m, but our ability to detect these bands for a particular target is subject to IWA constraints (see Section~\ref{sec:iwa_disc}).

\subsubsection{Biosignature False Positive Discrimination}

To study our ability to gather environmental context to identify false positives for biosignatures we also examined two worlds that may produce false positive biosignatures similar to the true biosignatures seen in our inhabited Archean-like and modern pre-industrial Earth-like planets. These false positive worlds include a prebiotic Earth-like world with high-volcanic outgassing from a more reducing mantle, which produces a false positive for the \ce{CH4}/\ce{CO2} disequilibrium pair \citep{Meadows2023}, and a habitable zone world with a dense ocean-loss \ce{O2} atmosphere and volcanic outgassing that retained an ocean, and that produces a strong abiotic \ce{O2} signal \citep{Meadows2018-gh}.

Our results show that for reflected light observations, the false positive case for the \ce{CO2}/\ce{CH4} pair is more readily confirmed  than the \ce{O2} false positive case considered, which is instead much harder to detect than the true inhabited case. 
For the prebiotic Earth-like world, the \ce{CO2} was comparably detectable to the Archean-like Earth for the Proxima b case, and the \ce{CH4} was only a factor of $\sim5$ less detectable, constituting a plausible false positive.  However, the CO discriminant that would indicate vigorous volcanic outgassing from a more reducing mantle \citep{Krissansen-Totton2018-tj,Thompson2022} is detectable with only $\sim10$ hours of observing in the most optimistic case for the prebiotic Earth atmosphere modeled here, allowing us to confirm the likelihood that we are observing a  \ce{CO2}/\ce{CH4} false positive atmosphere relatively quickly.

In comparison, for the post-ocean-loss planet, we find that the 10 bars of \ce{O2} in its atmosphere are significantly less detectable than the 0.2 bars of photosynthetically-generated \ce{O2} in the pre-industrial Earth atmosphere, confirming similar results from \citep{Leung2020}.  In the case of the 0.76 $\mu$m band, 20 times more observing time would be needed to detect this \ce{O2} band in the ocean-loss vs inhabited case, and four orders of magnitude more observing time would be needed for the 1.27$\mu$m \ce{O2} band making it basically undetectable.  This perhaps counterintuitive behavior is due to strong absorption in the bands
%saturation  contains 10 bars of \ce{O2}, the bands become saturated, and there is also 
and additionally strong \ce{O2}-\ce{O2} collisionally-induced absorption at wavelength coinciding with the 1.27$\mu$m \ce{O2} band \citep{Leung2020} which suppresses the high-resolution structure of these bands.  This  makes \ce{O2} in the false positive atmosphere less detectable than that for the pre-industrial Earth modern photosynthetic biosphere, and reduces the likelihood that a false positive is detected, but if it is, the much stronger suppression of the 1.27$\mu$m band is diagnostic of a multi-bar \ce{O2} atmosphere. Alternatively, signal extraction techniques on lower resolution spectra may be better suited for detecting these saturated absorption features; however, we leave this analysis to a future study.   In addition, pressure-broadened  water bands in the 10 bar post-ocean-loss environment require $\sim 10$ more hours to detect than the $\sim 1$ hr needed for the Proxima b habitable environments in the optimistic C=$10^{-4}$ cases, and the highly-oxidizing post-ocean-loss atmosphere also has undetectable \ce{CH4}.  Therefore, even if the weaker signals from the ocean-loss atmosphere are detected, detection of the \ce{O2} 0.76$\mu$m band, but not 1.27$\mu$m band, plus lack of or severe depletion of \ce{CH4} and \ce{H2O} observed in a comparable exposure time, would make biological processes as the origin of the observed \ce{O2} much less likely.

\subsection{Recommendations for an Observing Protocol}\label{sec:observingprotocol}
%DISCRIMINATE TERRESTRIAL FROM SUB-NEPTUNE -> CH4 FIRST, OR OPTIMIZED IF YOU HAVE A EXTENDED WAVELENGTH RANGE. 
%IF NOTHING DETECTED, WATER NEXT, ESTABLISHES HABITABILITY AND SERVES AS FALSE POSITIVE INDICATOR FOR DESICCATED PHOTOCHEMICAL PRODUCTION OF O2...ESPECIALLY SINCE PROBING TO THE VERY NEAR SURFACE IN THIS BAND?
%THEN CH4, CO2, O2 (1.27UM FP DISCRIMINANT! LEUNG ET AL.) CAN ALL BE OBSERVED SIMULTANEOUSLY IN SAME AMOUNT OF OBSERVING TIME IF WAVELENGTH RANGE POSSIBLE, OTHERWISE, PICK TWO.. CO2/CH4 LONGEST LIVED AND LIKELY MORE PREVALENT (EVOLVES FIRST...LESS COMPLICATED THAN OP). BUT ALSO O2/CH4 COMPARABLE DETECTABILITY, MUCH EASIER THAN FOR A G DWARF WORLD (cite sam's paper?).  IF WE DETECT CO2/CH4, THEN WANT CO TO RULE OUT FALSE POSITIVES. FOR O2/CH4, POSITIVE DETECTION OF WATER VAPOR + 1.27UM O2 RULES OUT MASSIVE ATMOSPHERE.  BUT IF YOU DON'T DETECT CH4 (NOT GUARANTEED TO CONVENIENTLY EVOLVE) THEN NEED 1.27 DETECTION (<10 BARS) AND/OR ABUNDANCE DETERMINATION. 

Given the detectability of molecular features in terrestrial and sub-Neptune atmospheres, we have developed an observing protocol detailing our recommendations for characterizing terrestrial planets and searching for signs of habitability and life with the ELTs, with particular emphasis on Proxima Centauri b since this is likely the most amenable to thorough atmospheric characterization. %Most known nearby non-transiting potentially terrestrial exoplanets are discovered by RV observations, and the true mass of the planet is unknown due to the Msin(i) degeneracy. 
The first step in our observing protocol is to discriminate a terrestrial atmosphere from a sub-Neptune atmosphere via the detection of \ce{NH3} or other reduced and/or H-bearing species, such as \ce{CH4} and CO, which can be optimized if the detector has an extended wavelength range between approximately 0.7 and 1.7 $\mu$m--- if the planet is more sub-Neptune-like, the atmosphere will likely be highly detectable, and many species, may be accessible in a single night of observing, allowing us to rapidly and conclusively identify the atmosphere as that of a sub-Neptune. If \ce{NH3} or other reduced species are not readily detected, this makes a terrestrial nature more likely, and we recommend followup observations to search for \ce{H2O} since it is likely detectable in about an hour of observation if Proxima Centauri b has a terrestrial atmosphere. A detection of water vapor could help establish habitability and serve as a false positive indicator for a  photochemical production of \ce{O2}, especially because the \ce{H2O} band  probes almost the entire atmospheric column, down to the very near surface. 

Next, \ce{CH4}, \ce{CO2}, and \ce{O2} can all be observed simultaneously within about 10 hours of observation time if it is possible to observe a broad $0.7\sim1.7$ $\mu$m wavelength range--- otherwise, choosing to search for only two, either \ce{CO2}/\ce{CH4} or \ce{O2}/\ce{CH4}, could provide pathways for detecting biosignatures or ruling out false positive cases. Searching for \ce{CO2} and \ce{CH4},  which absorb in a similar wavelength range, may be convenient if the instrument only has a narrow NIR bandpass. 
If \ce{CH4} and \ce{CO2} are positively identified in an atmosphere, we recommend searching for CO next. A detection of CO is possible in $\sim10$ hours for Proxima Centauri b for only the prebiotic Earth scenario in this work, and could indicate that the \ce{CH4} is abiotic (volcanically-produced) and that this is a false positive atmosphere. Conversely, a non-detection of CO in $>10$ hours could help rule out the biosignature false positive for Proxima Centauri b. 

Searching for \ce{O2}, and the \ce{O2}/\ce{CH4} biosignature pair, may also be possible simultaneously, however this would require the broader $0.7\sim1.7$ $\mu$m wavelength coverage because the absorption bands span the visible and NIR, respectively. Nevertheless, this biosignature pair may be more accessible for M dwarf systems than for G-dwarf systems \citep{Gilbert2024} due to the enhanced buildup of \ce{CH4} in M dwarf planet atmospheres \citep{Segura2005}. If \ce{O2} is detected in the visible, and not in the NIR, this could indicate that the atmosphere is an uninhabited ocean loss type world \citep{Leung2020}, and we find that a non-detection of \ce{CH4} and weak detection of \ce{H2O} would support this interpretation. Conversely, a strong detection of \ce{H2O} and the \ce{O2} 1.27 $\mu$m band rules out a massive \ce{O2} atmosphere; however, determining the absolute abundance of \ce{O2} would allow for a more definitive interpretation. We note also that instrumental contrast will in general be better at longer wavelengths since the contrast limited by wavefront errors scales as the inverse square of the wavelength. For example, in the case of the detectable \ce{O2} bands, this may result in up to 2.8x better contrast at 1.27 $\mu$m than at 0.76 $\mu$m.

\subsection{IWA limitations}\label{sec:iwa_disc}
% early vs. late type... early type probably good out to 5-10 pc because of HZ moving for these stars
%% maybe a couple targets... Wolf 1061 c and GJ 273 b
%% late type suffers because HZ is closer... PCb best target... 3+ pc you start to lose molecules redder than 1 um

% so, there are tradeoffs... early type the HZ is accessible for farther distances, but you have to contend with more stellar noise. Late type are amenable to short observations, but HZ is close to star and thus you have to have very close target... probably PCb is the best one, but you might be able to get O2 and maybe water for targets at ~3 pc (list targets here) 
% 
Despite the large diameters of the ELTs, which help improve the IWA ($\propto \lambda/D$) we can achieve, characterizing planets in the habitable zone of M dwarf stars will remain challenging. The habitable zones of these systems will be close to the host stars \citep[e.g.][]{Kopparapu2013-wm}, with typical angular separations on the order of 10--100 mas (Figure~\ref{fig:ang_sep}) for even the closest targets. While the angular separation is on the larger end of that range for early type systems (M0V--M3V), we may only be able to access habitable zone planets in wavelengths out to $\sim2 \mu$m for habitable zone planets orbiting early type targets $< \sim8-10$ pc away. Conversely, our ability to detect and characterize planets in the habitable zones of later type M dwarfs deteriorates rapidly with distance to the system due to the very small angular separations (10s of mas) between star and planet. Our analysis suggests that Proxima Centauri b is the only known potentially terrestrial HZ target for which we can access the full 0.5--2 $\mu$m wavelength range with the 39 m E-ELT. Beyond Proxima Centauri b, observing at wavelengths longward of $\sim 1 \mu$m will become challenging for even the closest known habitable zone planets orbiting mid-to-late M dwarfs, which include GJ 1061 \citep{Dreizler2020-ii}, Teegarden's Star \citep{Zechmeister2019-pa}, GJ 1002 \citep{Suarez_Mascareno2022-hd}, Wolf 1061 c \citep{wright2016three}, and GJ 273 b \citep{astudillo2017harps}. These IWA constraints will in turn make it difficult to detect spectral features from \ce{CH4}, \ce{CO2}, and \ce{CO}. This restricted wavelength range will significantly affect our ability to search for biosignatures, gain environmental context for an \ce{O2} detection, or rule out biosignature false positives. Accessibility to a broader range of molecules and multiple bands of the came molecule can be improved if the instrumental IWA can be refined; however, we acknowledge the inherent technical challenges this would require, and leave this as an open question for future instrumental design studies.

\subsection{Observing Logistics}
Here we have given detectability in total effective observing hours, but real-world consideration of observatory overheads, scheduling, telluric subtraction and weather, mean that hours of observing may take days to years of real-world time to complete. We use a simplistic parameterization to model  detectability as a function of time, and we do not include time lost due to observatory overhead such as slew and readout time, or other weather-related phenomena that can reduce the total available observation time on a given night. Additionally, we have assumed that all observations in this study occur at or near quadrature, which will not always be the case for real observations because of planet phase variability night-to-night, and from the beginning to the end of the night, especially for these M dwarf planets where the orbital period is measured in days. Removing telluric lines will also be a major challenge to overcome for real observations, and our results assume that we can model out the telluric lines. While we make this assumption for our work, there has been significant progress toward understanding how to optimally remove telluric lines \citep[e.g.][]{Cheverall2024}. Additionally, seasonal observability can limit us to certain optimal observing windows. Therefore, our results reflect the most optimistic observational conditions.

%\subsubsection{Nearby Targets}\label{sec:disc_realtargets}

If we take these realistic observing considerations into account then the hours required to detect molecules transforms into days and years.  We use an estimate for the total number of nights available to observe Proxima Centauri b calculated by \citet{Lovis2017-rr}: considering the contribution of weather, seasonal observability, and planetary phase, \citet{Lovis2017-rr} estimate that $\sim 20$ nights are available to observe Proxima Centauri b per year. 
%We note $\sim 20$ nights/year is likely a lower limit for 30 m class telescopes because of the $\lambda/D$ advantage gained with a larger telescope. Nevertheless, using this estimate and assuming a 30 year ELT operation lifetime, we will likely have at least 600 nights available to observe Proxima Centauri b over 30 years. 
If we assume 8 hr can be observed in a given night, Figure~\ref{fig:case_studies} suggests that many molecular targets will likely require less than a few years of operation time with the ELTs. Proxima b may be characterized within days to weeks, and targets farther away will likely take weeks to years, and may be limited by the instrumental IWA (see previous discussion).  Following the suggestion of \citet{Hardegree-Ullman2023}, an observing strategy that uses simultaneous observations of a target with more than one ELT could significantly reduce the time to detection, and maximize the molecular features that are accessible.

\begin{deluxetable*}{cc|cccc}\label{tab:pcb_compare}
\tablecaption{Configurations for Figure~\ref{fig:pcb_compare} detectability comparison}
\tablewidth{0pt}
\tablehead{
\colhead{Reference} & \colhead{Figure~\ref{fig:pcb_compare} panel} & \colhead{Telescope} & \colhead{Resolution} & \colhead{Contrast} & \colhead{Instrument/Background Noise}
}
\startdata
\citet{Lovis2017-rr} & a & VLT (8.2m) & 220,000 & $10^{-4}$ & On \\
\citet{Vaughan2024} & b & ELT (39m) & 17,385  & $10^{-4}$ & On \\
\citet{Hawker2019-lt} & c & ELT (39m) & 150,000 & $10^{-4}$ & Off \\
\citet{Wang2017} & d & TMT (30m) & 100,000 & $10^{-6}$ & On \\
\citet{Zhang2023} & e & ELT (39m) & 1,000 & $10^{-3}$ & On \\
\enddata
\end{deluxetable*}

\begin{figure*}
    \centering
    \includegraphics{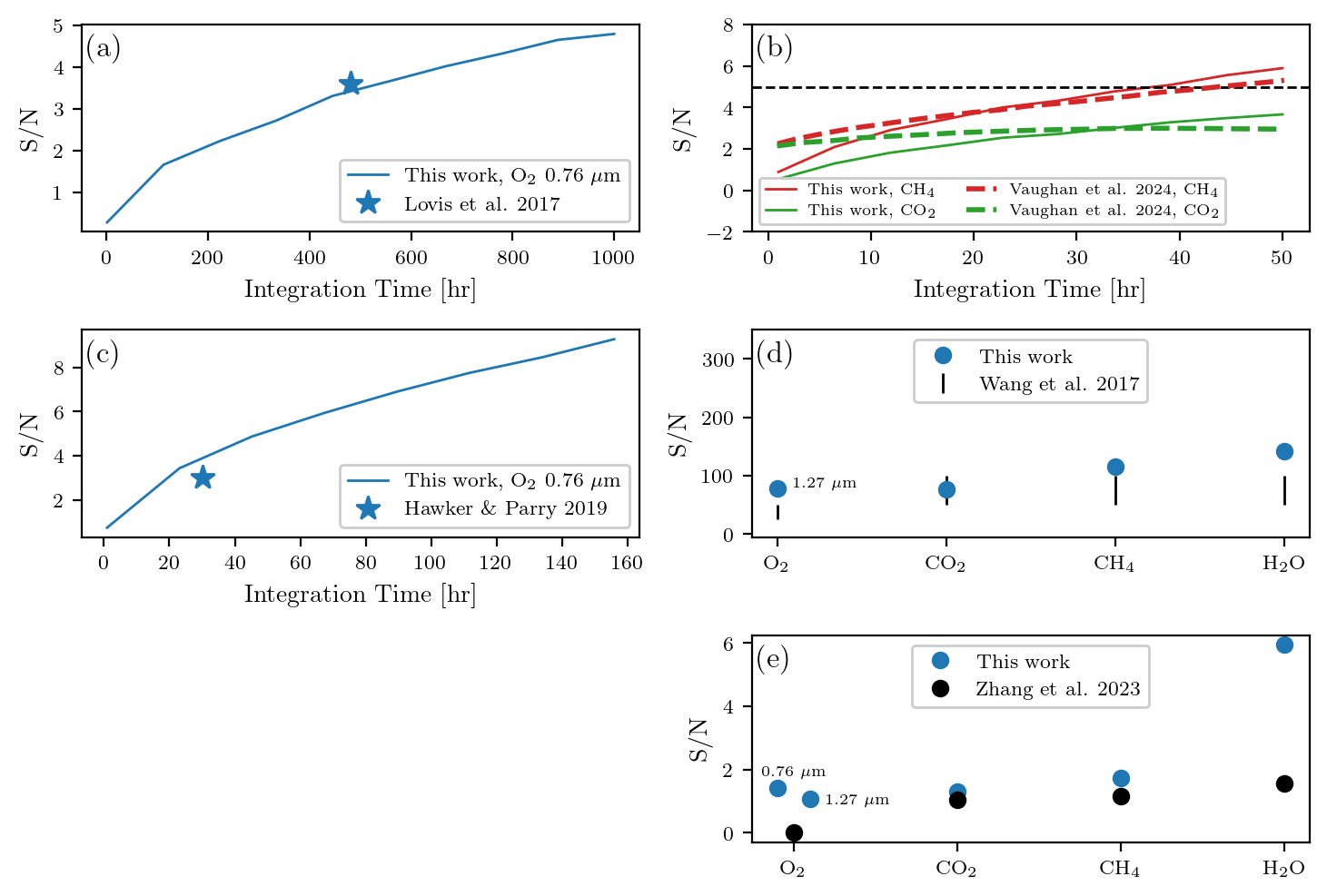}
    \caption{Comparison with previous work investigating the detectability of molecular features in an Earth-like atmosphere of Proxima Centauri b. Panels a, b, c, d, and e correspond to comparisons with \citet{Lovis2017-rr}, \citet{Vaughan2024}, \citet{Hawker2019-lt}, \citet{Wang2017}, and  \citet{Zhang2023}, respectively. The studies we compare with use a variety of observing modes and configurations, which we attempt to control for in our comparisons by using the same or similar input parameters, and we list our assumptions for observing conditions and configurations in Table~\ref{tab:pcb_compare}. Our results are largely consistent with these other studies, with the exception of more optimistic \ce{O2} and \ce{H2O} detection when compared to \citet{Zhang2023}.}% The comparison with \citet{Lovis2017-rr} (panel a) uses a VLT configuration, and we present detectability as a function of time that is consistent with their \ce{O2} result. We present credible detectability regions for \ce{O2}, \ce{CO2}, \ce{CH4}, and \ce{H2O} estimated from \citet{Wang2017} in panel b, where the black circles and arrows are lower and upper limits, respectively, and our results generally fall within these regions. In panel c we present a shaded region representing the region of detectability for the \ce{O2} A-band from \citet{Hawker2019-lt}, and our estimates are plotted as a curve which intersects with the \citet{Hawker2019-lt} result. Our comparison to \citet{Zhang2023} in panel d shows consistency in the \ce{CH4} and \ce{CO2} detectability, however our work is more optimistic for \ce{O2} and \ce{H2O}. Finally, panel e presents a comparison of \ce{CH4} and \ce{CO2} detectability with \citet{Vaughan2024}, where we show consistency in the trends.}
    \label{fig:pcb_compare}
\end{figure*}

\subsection{Comparison with other work}
Here we compare the results of this study to previous work that predicts the performance of current and future ground-based observatories to characterize terrestrial exoplanets in both reflected light and transit transmission observations.

\subsubsection{Reflected light observations}

The results of this study are largely consistent with previous work presenting detectability estimates for molecules in an Earth-like atmosphere on Proxima Centauri b. To optimally compare our work to previous work, we simulate molecular detectability by running our detectability pipeline customized with the same or similar observing conditions and modes used in \citet{Lovis2017-rr}, \citet{Wang2017}, \citet{Hawker2019-lt}, \citet{Zhang2023}, and \citet{Vaughan2024}. The observing configurations we assume include custom configurations of telescope size, spectral resolution, coronagraphic contrast and instrument and background noise, and are documented in Table~\ref{tab:pcb_compare}, and the results of our comparison are plotted in Figure~\ref{fig:pcb_compare}. The panels in Figure~\ref{fig:pcb_compare} represent comparisons to the works of \citet{Lovis2017-rr}, \citet{Vaughan2024}, \citet{Hawker2019-lt}, \citet{Wang2017},  and \citet{Zhang2023} (panels a, b, c, d, and e, respectively). We note however that our comparison input spectra were not standardized for abundances with published results, and so discrepancies in the more photochemically-active or variable molecules, such as methane and water vapor, is to be expected. 

\citet{Lovis2017-rr} simulate VLT observations of Proxima b, and we find that similar simulations using our detectability pipeline are consistent with their \ce{O2} result (see panel a of Figure~\ref{fig:pcb_compare}), but our work is more optimistic for \ce{H2O}. \citet{Lovis2017-rr} estimate that a  3.6$\sigma$ detection of \ce{O2} would require 480 hr of observing, and we match this estimate to within 10\%. However, we estimate that it will take a factor of $\sim10$ less time ($\sim 50$ hr vs. $\sim480$ hr) to detect \ce{H2O} than \citet{Lovis2017-rr} predict because they test only \ce{H2O} bands in the 0.6--0.7 $\mu$m region, where \ce{H2O} absorption is much weaker than in the 0.9 $\mu$m region where our estimates are made. 

\citet{Vaughan2024} simulate ELT HARMONI observations of Proxima Centauri b and target the molecular species \ce{CO2} and \ce{CH4}, and we find that an analogous observational configuration using our pipeline is consistent with their results. While \citet{Vaughan2024} do not explicitly state their effective contrast, we adopt their focal plane mask transmission value of $10^{-4}$ for a crude comparison. Because of this assumption, our results are likely on the optimistic side and our quoted exposure times can be thought of as lower limits.  In panel b of Figure~\ref{fig:pcb_compare}, we present this comparison showing very similar trends for both \ce{CO2} and \ce{CH4}. The results of \citet{Vaughan2024} indicate the possibility of some systematics that affect the shape of the curves, which our model does not reproduce--- \citet{Vaughan2024} use a highly sophisticated HARMONI noise model, which includes components that can introduce instrument-specific systematics, and this is neglected in our model. 

\citet{Hawker2019-lt} present a range of estimates for the \ce{O2} A-band detectability in the atmosphere of Proxima Centauri b using the ELT, and  our estimates are within these ranges when using a similar observing configuration. In panel c of Figure~\ref{fig:pcb_compare}, we plot the estimate of \citet{Hawker2019-lt}, where \ce{O2} is detectable at 3 $\sigma$. Our results are plotted as a curve of S/N as a function of integration time, which is within $\sim 30\%$ of the \citet{Hawker2019-lt} estimate.

\citet{Wang2017} simulate 100 hr observations of Proxima Centauri using a NIR detector on a 30 m telescope, and we find that our results are roughly consistent with their predictions for \ce{O2}, \ce{CO2}, \ce{CH4}, and \ce{H2O} detectability (see panel d of Figure~\ref{fig:pcb_compare}). Since precise values for molecular detectability were not presented  in \citet{Wang2017}, we  compare to detectability limits estimated from their figures. Since \citet{Wang2017} only simulate NIR observations, we only include an estimate for the 1.27 $\mu$m \ce{O2} band in our comparison. The black markers in panel d of Figure~\ref{fig:pcb_compare} represent an estimated lower bound on S/N gained from a 100 hr observation using a 30 m telescope, and the top of the arrows represent an upper bound. We find that our results are consistent with the credible detectability regions estimated from \citet{Wang2017}.

\citet{Zhang2023} simulate medium resolution (R=1,000) observations of Proxima Centauri b using the ELT, and using a similar configuration, we find that our SNR results are up to factors of up to 3 higher than \citet{Zhang2023}. Panel e of Figure~\ref{fig:pcb_compare} shows this comparison. Because \citet{Zhang2023} simulate a broad wavelength range that covers both the 0.76 $\mu$m and 1.27 $\mu$m \ce{O2} bands, we include estimates for both these bands in Figure~\ref{fig:pcb_compare}. Our results are most consistent for \ce{CH4} and \ce{CO2}, but we find major discrepancies when comparing \ce{O2} and \ce{H2O} detectability. \citet{Zhang2023} report that \ce{O2} is undetectable in 10 hrs of observing, while our results show that \ce{O2} is detectable using their observational configuration. We also report that \ce{H2O} is a factor of 3 more detectable than \citet{Zhang2023}. These discrepancies may be due to differences in S/N calculations: our detectability pipeline performs a cross-correlation analysis for many realizations of noise for simulated observations, while the S/N calculations of \citet{Zhang2023} rely on the analytical S/N equation, but identifying a precise cause is challenging without comparing input planetary spectra and analysis pipelines in depth. Furthermore, we note that telluric and stellar line subtraction may be extremely challenging at the resolution  used in \citet{Zhang2023} (R=1,000), as the telluric removal and cross-correlation techniques rely on high-resolution observations to correct for these effects.

We also confirm the results of \citet{Leung2020}, who investigated the relative detectability of \ce{O2} bands in scenarios with thick post-ocean-loss \ce{O2} atmospheres. They and we both find that suppression is expected in the 1.27 $\mu$m \ce{O2} band due \ce{O2}--\ce{O2} collisionally induced absorption, while the 0.76 $\mu$m A-band should remain more detectable. This mechanism is an \ce{O2} false positive discriminant. While \citet{Leung2020} did not simulate observations with realistic noise, we can confirm that this false positive discriminant is still valid, and possibly detectable, using our more realistic observation simulation and analysis pipeline. We indeed find that the detectability of the 1.27 $\mu$m \ce{O2} band is suppressed compared to the A-band, which indicates the dense post-ocean-loss \ce{O2} biosignature false positive atmosphere. Further evidence for this case could be gathered by attempting to retrieve the abundance of \ce{O2}, however we leave this to future work.

\subsubsection{Transit transmission observations}

\begin{figure}
    \centering
    \includegraphics{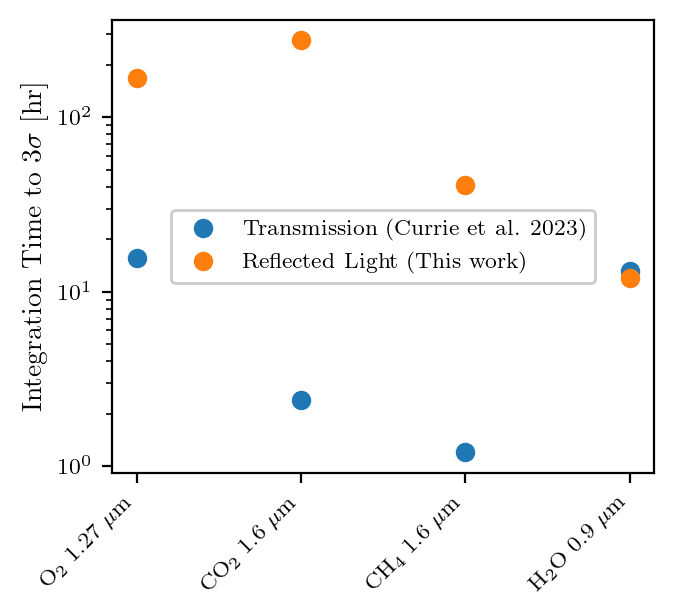}
    \caption{Integration time to detect \ce{O2}, \ce{CO2}, \ce{CH4}, and \ce{H2O} in a pre-industrial Earth-like atmosphere orbiting an M6V host star 5 pc away using both transmission and reflected light (C=$10^{-4}$) techniques, both with R=100,000. The transit transmission technique is overall more efficient for detecting \ce{O2}, \ce{CO2}, and \ce{CH4}, but these techniques yield comparable results for \ce{H2O} because reflected light observations are able to probe deeper into the atmosphere than in transmission and can detect the more abundance near-surface water. We emphasize that there is likely no transiting HZ terrestrial planet at 5 pc, and this figure should be only used as a reference for comparing differences in molecular detectability between the transmission and reflected light spectroscopy techniques.}
    \label{fig:trans_refl_compare}
\end{figure}

In this work, we considered several of the atmospheres used as input for the transit transmission calculations of \citet{Currie2023}, allowing us to directly assess the benefits and complementarity when observing equidistant targets in reflected light vs. transmission. Observing in reflected light allows us to sample the atmospheric column down to the surface or near surface of a planet, thereby enhancing sensitivity to gases that may be confined to the near-surface, such as \ce{H2O}. On the other hand, transit transmission observations are sensitive primarily to the upper troposphere and stratosphere of a terrestrial planet, and are not as susceptible to molecular band saturation as in reflected light. We also note the practical logistics of observing planets in transmission vs. reflected light can affect the total real-world time to molecular detections. For example, transits only occur once per orbit, while observing in reflected light may not be as limited in phase space, and there may be more opportunities to obtain reflected light observations; however, reflected light observations are subject to IWA considerations. Below, we compare the molecular detectability for reflected light (this study) to those we previously calculated for transit transmission \citep{Currie2023}.

We find that direct imaging targets require an additional step to rule out the planet being a sub-Neptune, as typically a lower limit on the mass is all that is known, compared to usually well-known radii and radius/density relations for transiting exoplanets. However, this additional step will likely only take a handful of observing hours per planet, and so is relatively inexpensive. 

Overall the transit transmission technique is more sensitive to most molecules in terrestrial planet atmospheres when compared to the reflected light technique, for equidistant targets, with the exception of water vapor, which is equally as detectable with the two techniques. Figure~\ref{fig:trans_refl_compare}  presents a comparison between the detectability of molecules in a hypothetical pre-industrial Earth-like atmosphere orbiting an M6V star at 5 pc away. We find that significantly more time is required to detect \ce{O2}, \ce{CO2}, and \ce{CH4} in reflected light for such a target than for transmission observations. To obtain a 3$\sigma$ detection for \ce{O2}, \ce{CO2}, and \ce{CH4}, an observer would need to integrate for approximately 13/168, 2/276, and 1/41 hours for transmission/reflected light, respectively. Transmission observations require $\sim10-100$x less observation time than reflected light observations to detect \ce{O2}, \ce{CO2}, and \ce{CH4} for a planet at 5 pc away. \ce{H2O} does not follow this trend for reflected light because reflected light observations are able to probe deeper into the atmosphere to reach the near-surface water, requiring nearly the same amount of observing time (12 hours for reflected light as opposed to 11 hours in transmission). 

However, we note these comparisons neglect to emphasize several observational and technological caveats.  The total observation time for transiting planets will be limited by the number of times we are able to observe a transit as well as the transit duration  itself. Furthermore, the S/N of the observation is limited by the shot noise of the entire stellar flux. Transiting planet observations may therefore be infrequent and have a strong stellar noise component. On the other hand, targets observed in reflected light may be obtained with higher frequency because there are more opportunities to observe them: targets in reflected light are optimally observed at and/or around quadrature (two quadratures per orbit), and are not limited to only the $\sim$hour long opportunities to observe once per orbit as for transiting targets.  Therefore, obtaining longer integrated exposure times for reflected light targets may be more feasible.   

We also note that because of the drop-off in reflected light sensitivity with planetary system distance, there will likely be very few habitable zone exoplanetary targets for which both transit and reflected light observations will be possible, unless a new transiting M dwarf HZ planet is discovered within 5 pc.  Overall, the \ce{CO2}/\ce{CH4} and \ce{O2}/\ce{CH4} biosignature disequilibrium pairs may be more accessible in transmission than in reflected light for a planet where both observational techniques are viable, but transiting terrestrial targets closer than 12 pc are unlikely. We are likely to have more and closer targets targets amenable to reflected light studies than transiting targets, and several direct imaging targets within 5pc are already known. 

If we compare instead the two best targets with each technique---TRAPPIST-1 e (at 12 pc) for transmission and Proxima Centauri b (at 1.3 pc) for direct imaging---we find that the reflected light technique allows us to access more molecules with fewer observational resources. \ce{O2} is more detectable for Proxima Centauri b, requiring at minimum 10 hours, while TRAPPIST-1 e may require 60--100 hours. \ce{CO2} and \ce{CH4} will likely require a similar time investment for Proxima Centauri b and TRAPPIST-1 e, requiring $\sim5$ and $\sim10$ hours for each, respectively. However, we note that the \ce{CO2}/\ce{CH4} biosignature pair is likely readily detectable for both of these targets, but the \ce{O2}/\ce{CH4} pair is likely best detected for non-transiting planets in reflected light. Finally, \ce{H2O} will likely be much more detectable for Proxima Centauri b, requiring nearly 100x less observation time than for TRAPPIST-1 e. Overall, characterizing the nearest non-transiting target Proxima Centauri b will likely require fewer resources than the nearest transiting target, TRAPPIST-1 e.

\section{Conclusions}\label{sec:conclusions}

In this work, we upgraded the existing SPECTR ELT detectability pipeline to include functionality for high spectral resolution observations of terrestrial exoplanets in reflected light, and applied SPECTR to estimate the detectability of molecular features in simulated observations of five exoplanet atmosphere types, including pre-industrial Earth-like, Archean Earth-like, prebiotic Earth-like, an ocean loss scenario, and a sub-Neptune atmosphere.
We find that \ce{O2}, \ce{CO2}, \ce{CH4}, and \ce{H2O} may be detectable for nearby non-transiting targets in as little as a few hours of observing with an ELT for the most optimistic cases, and we may be able to readily search for signs of habitability and life, or rule out habitable but uninhabited environments that include biosignature false positive gases, and sub-Neptune like atmospheres. If Proxima Centauri b, which is the prime target for highly detailed atmospheric characterization in the ELT era, is a sub-Neptune, this will be readily apparent via detection of reducing molecules like \ce{NH3} which is feasible in approximately an hour of observing with the ELT. We also show that \ce{CO2} and \ce{H2O} are comparatively detectable in sub-Neptune and terrestrial atmospheres, and so are poor discriminants for terrestrial worlds. For a terrestrial Proxima Centauri b, two biosignature disequilibrium pairs are potentially accessible in $\sim 10$ hours  with broad wavelength coverage (0.5--2.0 $\mu$m), including \ce{O2}/\ce{CH4} and the \ce{CO2}/\ce{CH4} pairs. We may be able to discriminate biosignature false positives via detection of CO and failure to detect \ce{H2O} or \ce{CH4} in a comparable amount of observation time. 

Multiple other targets farther away than PCb, but within 5 pc away, will likely be amenable to atmospheric characterization using this technique at the cost of approximately an order of magnitude or more observation time, and molecular bands $> 1 \mu$m may be challenging to observe due to IWA limitations, and  as a result these atmospheres may be characterized less thoroughly. We compared the efficacy of transmission vs reflected light observations for characterizing terrestrial exoplanets, and while transmission observations were more sensitive to molecular bands for a hypothetical equidistant target, in reality direct imaging has more and closer targets to study.  While terrestrial exoplanet characterization is inherently challenging, the ELT era will likely offer the first opportunities to study the atmospheres of non-transiting terrestrial exoplanet targets, and search for signs of habitability and life on our nearest exoplanetary neighbors.

\section{Acknowledgements}
The authors thank the anonymous reviewers for their thoughtful comments and suggestions that greatly improved the quality of this manuscript. We also thank Andrew Lincowski and Drake Deming for our helpful discussions, and C. Evan Davis, who ran the original climate--photochemistry calculations for the the pre-industrial Earth and Archean Earth atmospheres used here. This work was performed by the Virtual Planetary Laboratory Team, a member of the NASA Nexus for Exoplanet System Science (NExSS), funded via the NASA Astrobiology Program grant No. 80NSSC18K0829. This work was also partially supported by the Astrobiology Center, Japan. The simulations in this work were facilitated though the use of advanced computational, storage, and networking infrastructure provided by the Hyak supercomputer system at the University of Washington.

\appendix

\section{Supplemental Spectra}\label{apx:spectra}
\begin{figure*}
    \centering
    \includegraphics[width=\linewidth]{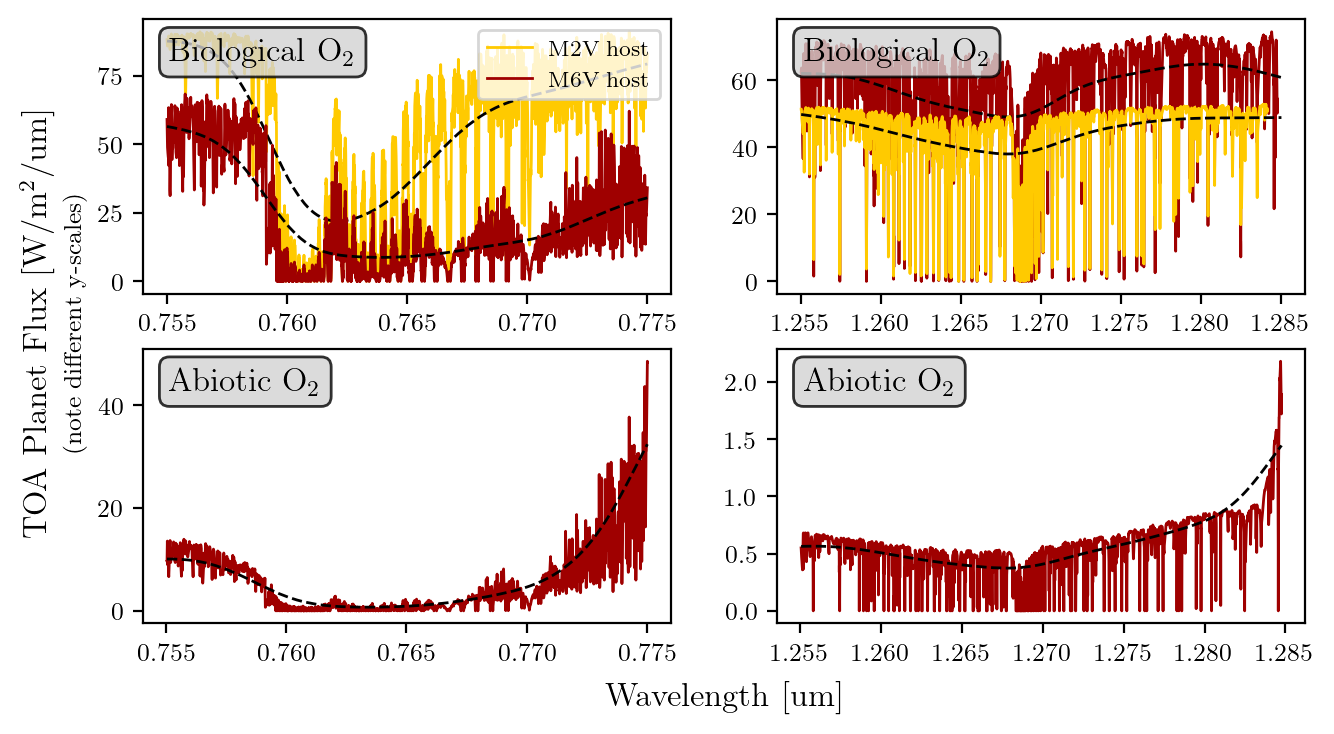}
    \caption{Examples of top of atmosphere (TOA) planetary flux spectral models for \ce{O2} biosignature and false positive cases. The left and right columns present the \ce{O2} 0.76 and 1.27 $\mu$m bands, respectively, for the biological \ce{O2} pre-industrial Earth-like case (upper) and abiotic \ce{O2} ocean loss case (lower). The yellow and red curves represent spectra that are self-consistent with M2V and M6V host stars, respectively. We plot low spectral resolution (R=500) curves over each high-resolution spectrum as black dashed curves. Note that we present the abiotic \ce{O2} with only a M6V host as an atmospheric model consistent with an M2V host was unavailable.  }
    \label{fig:spectra_o2}
\end{figure*}

\begin{figure*}
    \centering
    \includegraphics[width=\linewidth]{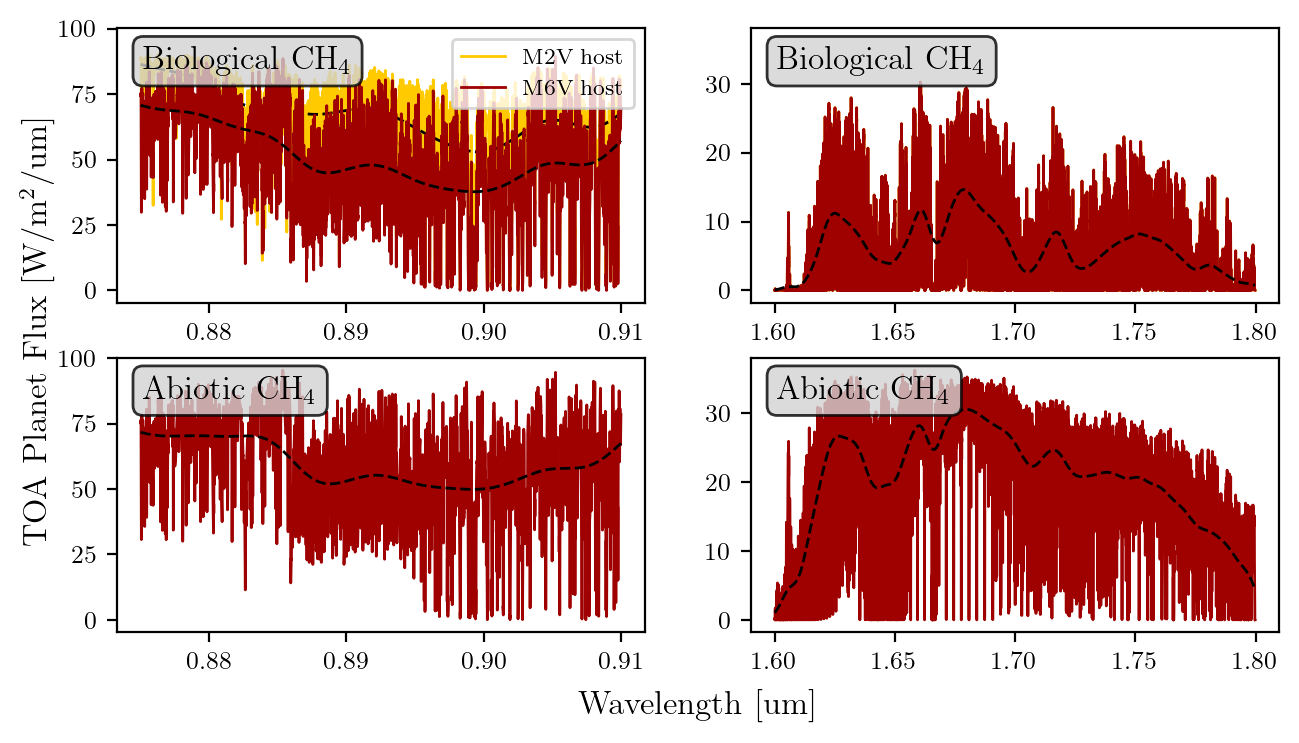}
    \caption{Examples of top of atmosphere (TOA) planetary flux spectral models for \ce{CH4} biosignature and false positive cases. The left and right columns present the \ce{CH4} 0.89 and 1.6 $\mu$m bands, respectively, for the biological \ce{CH4} Archean Earth-like case (upper) and abiotic \ce{O2} prebiotic Earth-like case (lower). The yellow and red curves represent spectra that are self-consistent with M2V and M6V host stars, respectively. We plot low spectral resolution (R=500) curves over each high-resolution spectrum as black dashed curves.  Note that we present the abiotic \ce{O2} with only a M6V host as an atmospheric model consistent with an M2V host was unavailable.  }
    \label{fig:spectra_ch4}
\end{figure*}

Here we include high-resolution spectra of top-of-atmosphere planetary flux models used as input for this work. Figures~\ref{fig:spectra_o2} and~\ref{fig:spectra_ch4} present spectra for both biosignature and biosignature false positive cases for the molecules \ce{O2} and \ce{CH4}, respectively, for M2V and M6V host stars. 

\section{Instrumental and Observatory Dependencies}\label{apx:dependencies}
We test the dependence of molecular detectability on three instrumental and observatory parameters to show expected trends in detectability dependence, including the distance to the exoplanetary system, stellar light suppression efficiency, and the total collecting area of the observatory. By testing how detectability scales with these parameters we can predict the first-order performance of the ELTs to characterize terrestrial exoplanet atmospheres in a range of observational scenarios. To illustrate these effects, we present detectability as a function of these three parameters for only pre-industrial Earth like atmospheres in Figures~\ref{fig:distance}, \ref{fig:contrast}, and~\ref{fig:observatory}. It is straightforward to calculate the theoretical dependence of integration time on these parameters: integration time to first order scales as distance squared ($d^2$), the inverse of the effective instrumental contrast ($C^{-1}$), and the inverse of the primary mirror diameter squared ($D^{-2}$).  Indeed, our results show these trends as the curves plotted in Figures~\ref{fig:distance}, \ref{fig:contrast}, and~\ref{fig:observatory}. These scalings hold for our assumption that our observations are to first order limited by only the photon noise of residual stellar light, and will break down as the planet/star contrast approaches the instrument contrast, or at very high spectral resolution as the detector noise becomes more dominant.  %Molecular detectability will ultimately be limited by the number of target photons we are able to collect on the detector, a prime motivation to construct the ELT class of observatories.  

\subsection{Distance Dependence}\label{sec:distance_dep}
\begin{figure}
    \centering
    \includegraphics{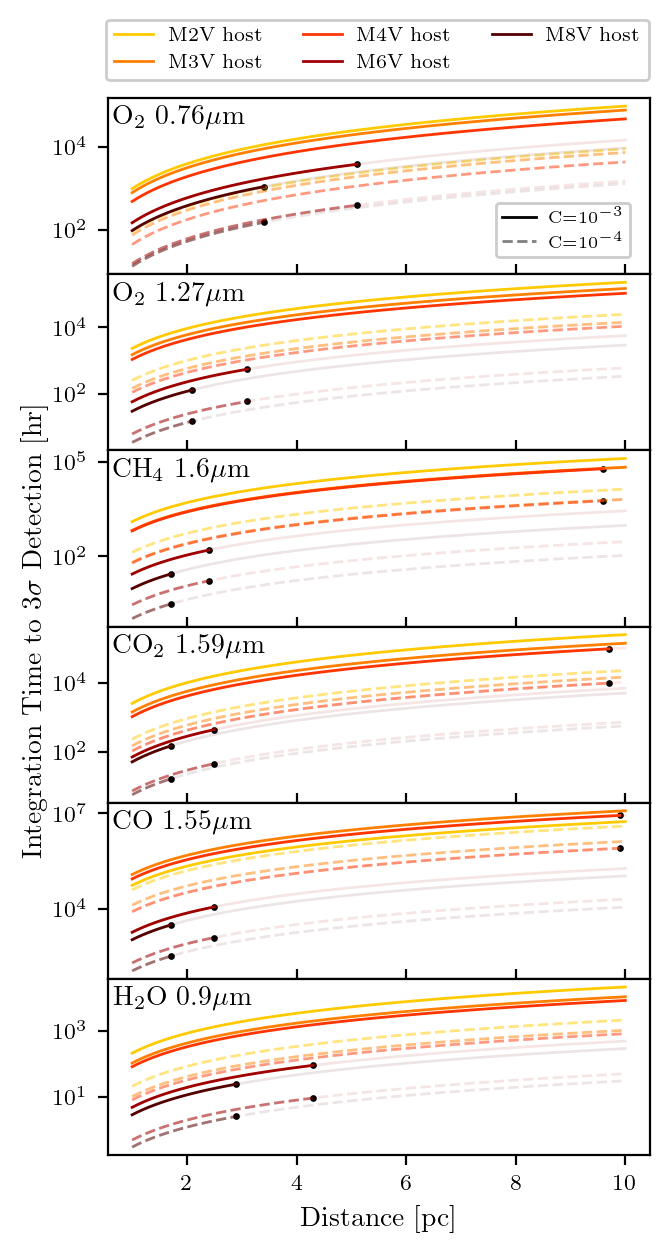}
    \caption{Time to a $3 \sigma$ detection as a function of distance to the system for pre-industrial Earth atmospheres orbiting M dwarf hosts with systemic RVs of 20 km/s, observed with the E-ELT with instrumentation capable of R=100,000 and C=$10^{-4}$, and with a total throughput of 10\%. The time to detection scales as $d^2$. For each curve, we plot a black marker where the planetary separation falls below our assumed 2 $\lambda/D$ IWA for a 39 m mirror--- targets with distances greater than this value will be challenging to observe. Later type targets are more affected by the IWA limitation, and Proxima Centauri b ($\sim$M6V, 1.3 pc away) will likely be the only nearby late type target for which all tested molecular bands are accessible.}
    %Early type targets beyond 5 pc away may not have readily accessible molecular features, but  late-type targets may be reasonably accessible for up to 10 pc.  For this plot, the effects of telescope inner working angle have been neglected.}
    \label{fig:distance}
\end{figure}

In Figure~\ref{fig:distance}, we plot the observing time required for a $3 \sigma$ detection of select molecular bands from Section~\ref{sec:molecular_detectability} as a function of distance. We assume these pre-industrial Earth-like atmospheres orbit M2V--M8V host stars with 20 km/s systemic RVs, and are observed using E-ELT instrumentation capable of R=100,000 and a contrast ratio (C) of $10^{-4}$. We note that the contrast used in these calculations is a simplistic assumption--- the true effective contrast of real instrumentation will depend strongly on the separation of the target. For each host star curve, we compute the time to detection for a grid of 20 evenly-spaced distance values between 1 and 10 pc, then fit these values to a curve $\propto d^2$, which is ultimately plotted in the panels. %As expected, the curves in Figure~\ref{fig:distance} suggest that time to detection scales as $ d^2$, where $d$ is distance to the system. 
In Figure~\ref{fig:distance}, we have additionally plotted the limiting distance where the separation of the planet and star falls within the inner working angle as a black marker for each curve. Targets at distances greater than this value will be challenging to image with our assumed IWA of 2$\lambda/D$. If a curve does not have a black marker, this means the limiting distance is $>10 $pc. See Section~\ref{sec:iwa_disc} for a discussion of impact of IWA on our ability to characterize habitable zone exoplanets. 
% For early-type hosts, it may be challenging to detect molecular features in planetary atmospheres beyond 5 pc away from Earth. Planets around late type hosts may be amenable to characterization beyond 5 pc, but obtaining multiple molecular species detections beyond 10 pc may be challenging, some requiring $>10^3$ hours. For example, at 5 pc, \ce{O2}, \ce{CH4}, and \ce{CO} require $>10^3$ hours and \ce{H2O} requires $>10^2$ hours for early type systems, while these molecules are all still accessible to late type systems, barring IWA constraints. At 10 pc, late type systems require $10^3$ hours for \ce{O2} (0.76 $\mu$m) and CO, $>10^2$ hours for \ce{O2} (1.27 $\mu$m), \ce{CH4}, and \ce{CO2}, and $>10$ hours for \ce{H2O}. However, late type hosts have habitable zones that are closer to the star, and IWA constraints may begin to limit our ability to resolve the planet signal for $>5$pc distant targets (Figure~\ref{fig:ang_sep}). 

\subsection{Stellar Suppression Performance}
% this weird flattening is due to sky background and instrument noise 

\begin{figure}
    \centering
    \includegraphics{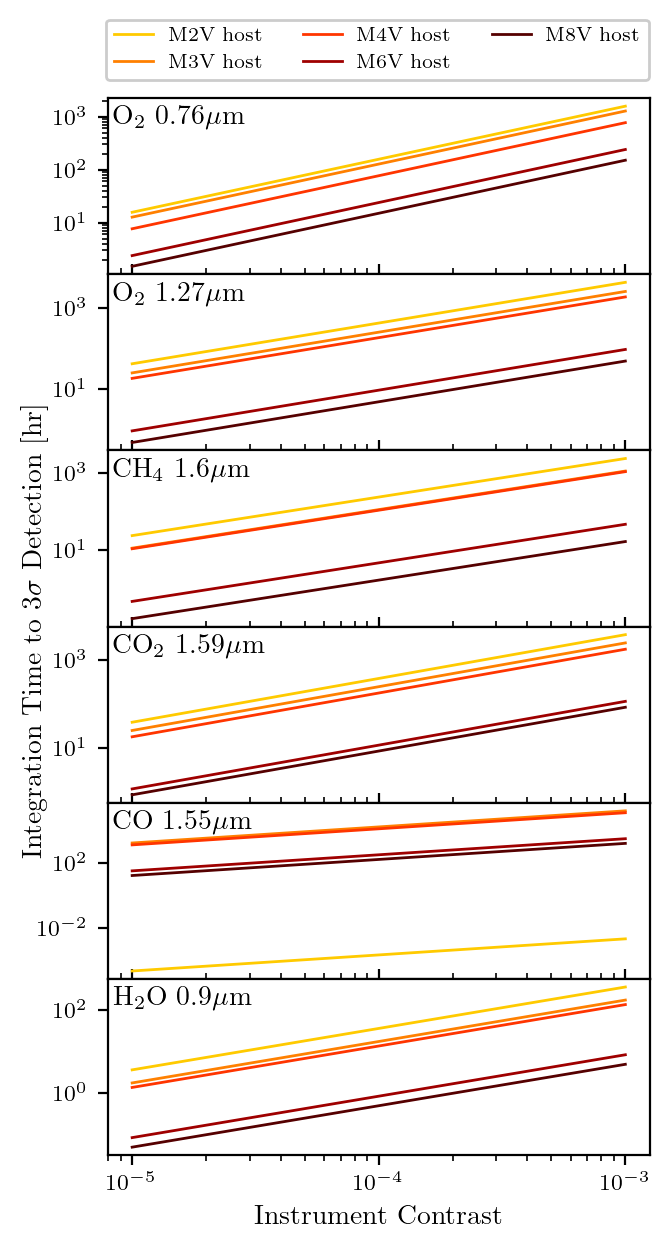}
    \caption{Time to a $3 \sigma$ detection as a function of instrumental contrast for pre-industrial Earth atmospheres orbiting M dwarf hosts for an E-ELT with instrumentation capable of spectral resolution R=100,000 and a total throughput of 10\%.  The M dwarf hosts are assumed to be 1.3 pc and have systemic RVs of 20 km/s. }
    \label{fig:contrast}
\end{figure}

We also explore the impact of instrument starlight suppression (contrast ratio) on exposure time needed to achieve a 3$\sigma$ detection, and plot exposure time as a function of contrast ratios spanning $10^{-5}$ and $10^{-3}$ in Figure~\ref{fig:contrast} for pre-industrial Earth-like atmospheres orbiting a range of M dwarf hosts.  
As expected, increasing the contrast ratio (smaller values) decreases the time required to reach the 3-$\sigma$ detection because this significantly reduces the stellar component of the noise budget. Improvements in time to detection are up to two orders of magnitude better for $10^{-5}$ as opposed to $10^{-3}$  contrasts.

\subsection{Observatory Dependence}
\begin{figure}
    \centering
    \includegraphics{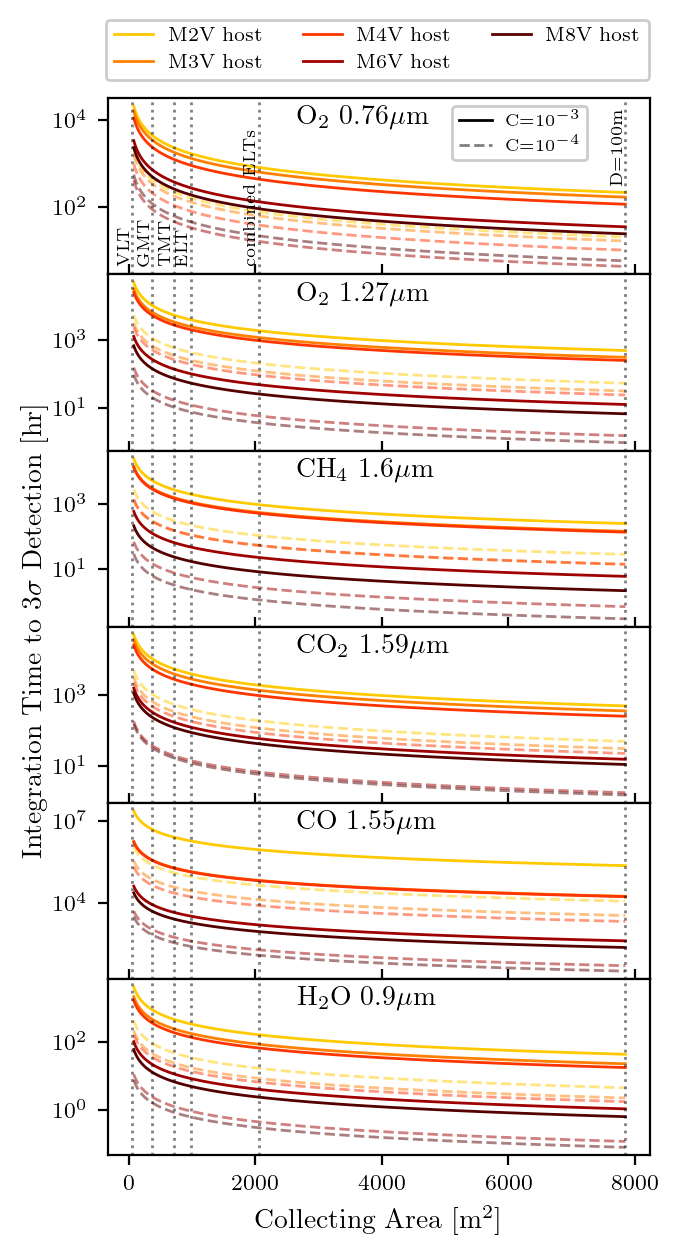}
    \caption{Time to a $3 \sigma$ detection as a function of telescope collecting area for pre-industrial Earth atmospheres orbiting M dwarf hosts 4.0 pc away with systemic RVs of 20 km/s, observed with with instrumentation capable of R=100,000 and a total throughput of 10\%. Solid and dashed lines represent assumed instrumental contrasts of C=$10^{-3}$ and C=$10^{-4}$, respectively. Current or planned telescopes are shown as dashed lines and labeled in the top panel.  As expected, larger apertures improve the time to detection in all cases. }
    \label{fig:observatory}
\end{figure}

Here, we simulate the detectability of select molecular features as a function of telescope collecting area, and in  Figure~\ref{fig:observatory} we report the time to $3 \sigma$ detection for pre-industrial Earth atmospheres orbiting a range of M dwarf hosts.
We compute the time to detection for a suite of 10 collecting areas ranging from the VLT (8.2 m diameter, 53 m$^2$ collecting area) to a hypothetical 100 m diameter ($\sim7900 \mathrm{m}^2$ collecting area) telescope \citep[e.g.][]{Dierickx2000}.  The diameters of current and future observatories, as well as a calculation for simplistically combining observations using all three planned ELTs, are shown as vertical lines in  Figure~\ref{fig:observatory}, and are labeled in the top panel. Increasing the collecting area from a VLT-sized telescope to the ELT class yields an improvement in time to detection by at least an order of magnitude in all cases. Combining ELT observations for a single target can decrease overall time to molecular detection by up to about an order of magnitude in some cases. 

\section{Results for GMT and TMT}\label{apx:GMGTMTres}
\begin{figure*}
    \centering
    \includegraphics{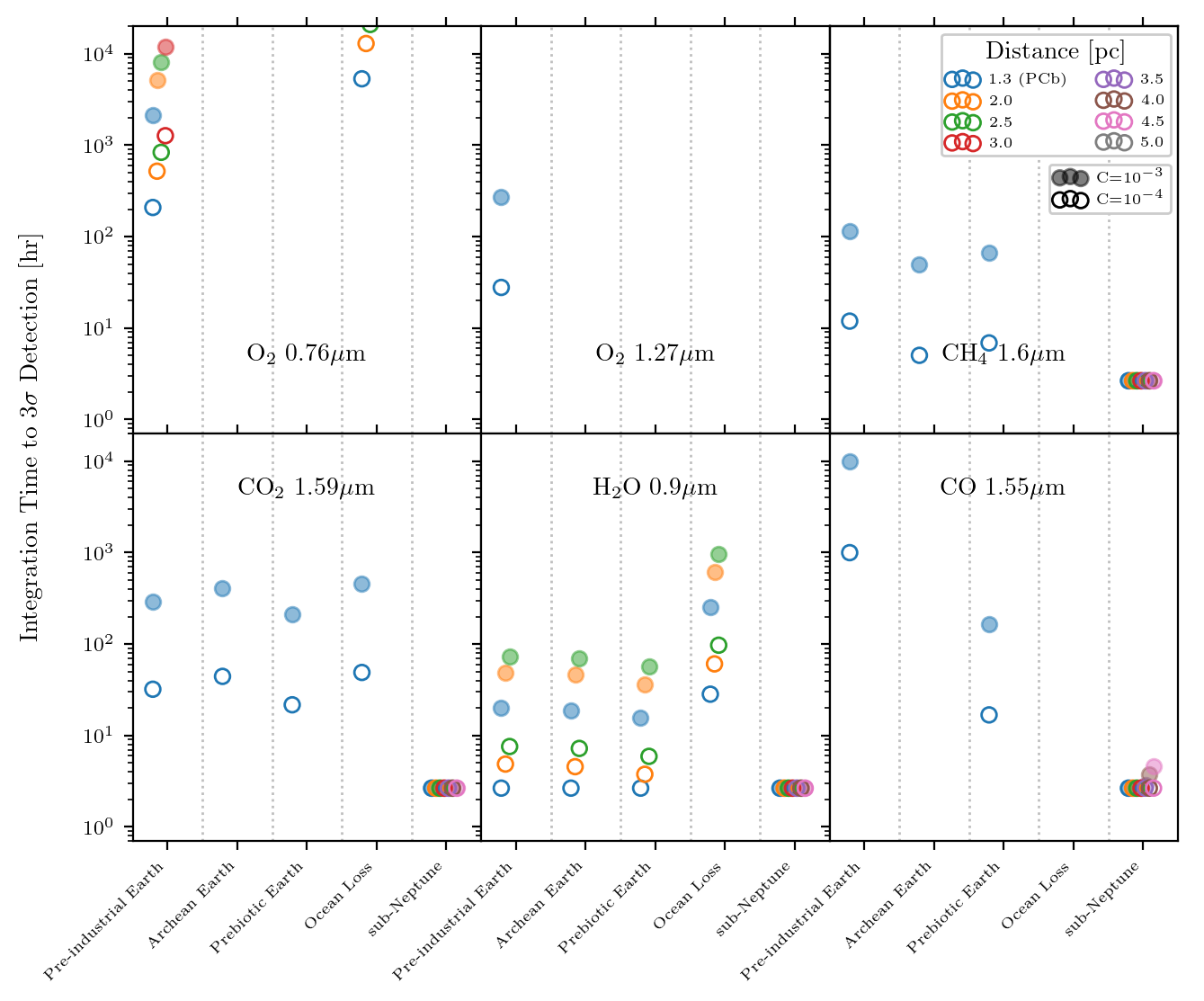}
    \caption{Same as Figure~\ref{fig:case_studies}, but for the GMT.}\label{fig:case_studies_GMT}
\end{figure*}

\begin{figure*}
    \centering
    \includegraphics{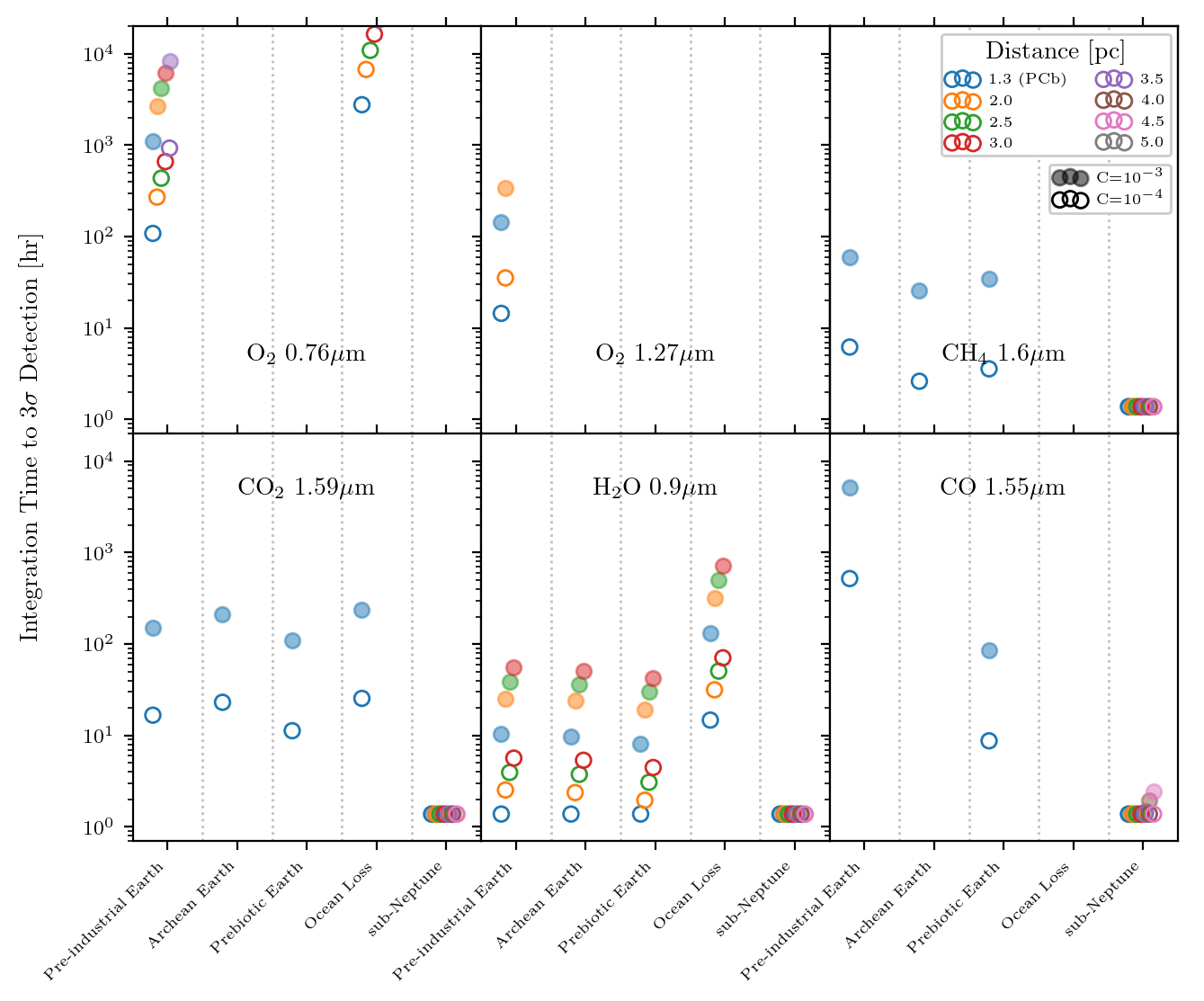}
    \caption{Same as Figure~\ref{fig:case_studies}, but for the TMT.}\label{fig:case_studies_TMT}
\end{figure*}

While the results discussed in this paper primarily focused on the capabilities of the European ELT because it will be the least limited by the IWA $\propto \lambda/D$ effect, here we also present selected results for the GMT and TMT similar. Similar to Figure~\ref{fig:case_studies}, we present mid-type (M6V) M dwarf target results covering the full range of atmosphere types we tested in this work. Ostensibly, our results for the GMT and TMT in Figures~\ref{fig:case_studies_GMT} and~\ref{fig:case_studies_TMT}, respectively, simply show the expected increase in effective observation time due to the reduction in collecting area. We use the same generic instrumentation setup outlined in Section~\ref{sec:tellurics} for all three ELTs. 

\bibliography{references}

\end{document}